\documentclass[]{aa}  
\usepackage{graphicx}
\usepackage{multirow}
\usepackage[table,xcdraw]{xcolor}
\usepackage{colortbl} 

\usepackage{txfonts}
\usepackage{colortbl}
\usepackage{hyperref}
\hypersetup{colorlinks=true,linkcolor=[rgb]{1.,0.2,0.2},citecolor=[rgb]{0.1,0.1,1.},filecolor=[rgb]{0.7,0.2,0.2},urlcolor=[rgb]{0.7,0.2,0.2}}

\usepackage{xcolor}
\definecolor{peru}{RGB}{205,133,63}
\definecolor{darkorange}{RGB}{255,140,0}
\definecolor{cornflowerblue}{RGB}{100,149,237}
\definecolor{slateblue}{RGB}{106,90,205}
\definecolor{salmon}{RGB}{250,128,114}
\definecolor{firebrick}{RGB}{178,34,34}
\definecolor{medianblue}{RGB}{0,0,205} 
\definecolor{brown}{RGB}{165,42,42}

\usepackage{pifont}
                                
\usepackage{hyperref}
\hypersetup{colorlinks=true,linkcolor=[rgb]{1.,0.2,0.2},citecolor=[rgb]{0.1,0.1,1.},filecolor=[rgb]{0.7,0.2,0.2},urlcolor=[rgb]{0.7,0.2,0.2}}

\newcommand{\Threep}{${+3\sigma}$}
\newcommand{\Twop}{${+2\sigma}$}
\newcommand{\Onep}{${1\sigma}$}
\newcommand{\Twom}{${-2\sigma}$}
\newcommand{\Threem}{${-3\sigma}$}
\newcommand{\MSG}{MS+\texttt{Grafting}}

\newcommand{\msun}{\,\mathrm{M_\odot}}

\newcommand{\lgbh}{\texttt{L-Galaxies}\textit{BH}} 
\newcommand{\lgalaxies}{\texttt{L-Galaxies}}

\begin{document}

\title{Overmassive and Undermassive Massive Black Holes:\\ The Role of Environment and Gravitational-Wave Recoils}
\titlerunning {Shaping the scaling relation: Environment and Gravitational-Wave Kicks}
%%%%%%%%%%%%%%%%%%%%%%%%%%%%%%%%%%%%%%%%
% Please do not include ORCIDs next to author names.
% Only ORCIDs authenticated by individual authors in EDP Sciences editorial system will be taken into account.
% ORCIDs included here will be removed.
%%%%%%%%%%%%%%%%%%%%%%%%%%%%%%%%%%%%%%%%

   \author{David Izquierdo-Villalba\inst{1}\fnmsep\thanks{dizquierdo@ice.csic.es}
        }

   \institute{$^1$ Institute of Space Sciences (ICE, CSIC), Campus UAB, Carrer de Magrans, E-08193 Barcelona, Spain\\
   }

   \date{Received ---, Acepted ---}

  \abstract{Understanding the connection between galaxy properties and their central massive black holes (MBHs) is key to unveiling their co-evolution. We use the {\tt L-Galaxies}-{\it BH} semi-analytical model and the {\tt Millennium} suite of simulations to investigate the physical origin of galaxies hosting overmassive and undermassive MBHs with respect to the $M_{\rm BH}$–$M_*$ relation, across stellar mass and cosmic time. We find that distinct evolutionary pathways drive different offsets from the scaling relation. Overmassive MBHs are primarily associated with galaxies that experienced enhanced merger history and secular activity. At $z\,{>}\,4$, this activity often leads to early, rapid MBH growth, frequently involving super-Eddington accretion episodes. At low redshift, a minority of overmassive systems ($20\%$) instead arise from environmental effects that reduce the stellar mass of the host, shifting galaxies above the relation without requiring additional MBH growth. Undermassive MBHs originate from two main channels. In massive galaxies, gravitational recoil following MBH mergers can eject the central MBH, temporarily leaving the galaxy without a nucleus. During this phase, MBHs coming from previous galaxy mergers can become the new central MBHs, but their masses remain below the expected ones from the scaling relation, as they never co-evolved with their new host galaxy. In low-mass galaxies ($M_*<10^9 M_\odot$), undermassive MBHs are more commonly linked to a quiescent evolutionary history, with limited mergers and weak secular processes that suppress an efficient MBH growth.  We therefore conclude that outliers of the $M_{\rm BH}-M_*$  do not arise from a single mechanism, but from the interplay between environmental effects, gravitational recoils, and diverse MBH fueling histories, whose relative importance varies with galaxy mass and redshift. }


   \maketitle

\section{Introduction}

Significant advancements in the observational study of Active Galactic Nuclei (AGN) have provided growing evidence that massive black holes (MBHs) with masses above $10^5\, \msun$ form naturally in galaxies and power AGN activity through gas accretion at their centres \citep{Schmidt1963, MerloniANDHeinz2008, Ueda2014, Hopkins2007, Aird2015}. Moreover, AGN demographic studies and dynamical analyses of stars and gas in the central regions of nearby galaxies have shown that a large fraction of massive galaxies in the local Universe host MBHs in their nuclei \citep{Genzel1987, Kormendy1988a, Dressler1988, Kormendy1992, Genzel1994, Peterson2004, Vestergaard2006}.

Over the past few decades, numerous studies have advanced our understanding of MBH formation and growth, revealing a series of tight correlations between MBH mass and key properties of their host galaxies. One of the first relations was reported by \citet{KormendyAndRichstone1995}, who found a connection between MBH mass and bulge luminosity. Shortly thereafter, \citet{Magorrian1998} established that MBH mass correlates with bulge stellar mass, an association later refined by \citet{MarconiandHunt2003} and \citet{HaringANDRix2004}. An even tighter correlation was identified by \citet{Ferrarese2000}, who showed that MBH mass scales strongly with the stellar velocity dispersion \citep[see also][]{Gebhardt2000}. Additional relations have been proposed, such as those linking MBH mass to the Sérsic index of the galaxy \citep{Graham2001} and to the total stellar mass of the host \citep{ReinesVolonteri2015,Erwin2012,Capuzzo2017,Harikane2023}. All these scaling relations carry significant physical implications, as they point to a close co-evolution between galaxy assembly and MBH growth. On one hand, they suggest that star formation (SF), bulge build-up, and gas accretion onto the MBH are interconnected processes \citep[see, e.g][]{SilkAndRees1998,Croton2006,Wild2010,Ishibashi2012,DeGraf2015,Carraro2020,Habouzit2019,Habouzit2021}. On the other hand, they imply that feedback from actively accreting MBHs can influence the baryonic evolution of the host galaxy, regulating SF and shaping its stellar mass build-up \citep[see, e.g.][]{DiMatteo2005,Croton2006,Croton2006a,DeGraf2017,MartinNavarro2018}.

The estimation of MBH masses used in the scaling relations has been commonly carried out using dynamical modelling of the motions of gas and stars in the galactic centre \citep[][]{KormendyAndHo2013}, reverberation mapping techniques \citep{Peterson2004,Lu2025}, or X-ray variability methods \citep[e.g.][]{McHardy2006}. In contrast, the galaxy properties such as stellar mass are typically inferred from mass-to-light ratios or from spectral energy distribution (SED) fitting \citep[][]{Conroy2013}. Although each of these approaches carries its own systematic and statistical uncertainties \citep[see, for instance,][]{VestergaardAndPeterson2006, Shen2013}, the resulting measurements consistently show that the established scaling relations exhibit an intrinsic scatter, with a population of overmassive MBHs (lying above the median relation) and undermassive MBHs (lying below it). Regarding the former population, overmassive MBHs have been observed across different cosmological epochs. Recent results at $z \,{>}\, 5$ from the James Webb Space Telescope (JWST) have unveiled a population of MBHs that are tentatively up to two orders of magnitude more massive than their local counterparts \citep[see e.g][]{Maiolino2023,Harikane2023,Kocevski2023,Pacucci2023,Matthee2024,Jones2025}. These discoveries have challenged the standard paradigm of MBH formation and evolution, suggesting that MBHs either originate from heavy seeds (${\sim}\,10^4\,\msun$) or grow from light seeds (${\sim}\,100\, \msun$) through episodic super-Eddington accretion \citep{Inayoshi2025, Bonoli2025}. Similar overmassive MBHs are also observed at lower redshifts ($z\,{\leq}\,3$), implying that such systems are not necessarily outliers in the context of the galaxy–MBH correlation but a natural outcome \citep[][]{Ferre_Mateu2021,Mezcua2023,Mezcua2024}. Concerning the population of undermassive MBHs, several studies have reported their presence in the local Universe \citep{Hu2008,KormendyAndHo2013,Savorgnan2016}, often associated with pseudobulge-dominated galaxies. Similar cases are observed in the AGN population \citep{GrahamANDLi2009}, suggesting that active, undermassive MBHs may be in a phase of rapid growth, potentially evolving toward the median scaling relation.

All the recent observational discoveries motivate the identification and theoretical characterisation of MBH outliers in the scaling relation, as they offer a powerful probe of the physical processes governing MBH growth and galaxy evolution. For instance, overmassive MBHs, which lie above the median $M_{\rm BH}\,{-}\,M_*$ relation, might indicate systems where black hole growth has proceeded more rapidly than the assembly of the host galaxy stellar mass \citep[see e.g,][]{Weller2023}. This accelerated growth may result from prolonged or highly efficient accretion episodes. Conversely, the population of undermassive MBHs, found below the median relation, may correspond to galaxies in which stellar mass assembly has overpassed the MBH growth rate (either through efficient secular or external processes), systems where MBH growth is still in an early phase or cases where the MBH assembly mechanism is hindered by different processes that limit gas inflow \citep[e.g.][]{VolonteriNatarajan2009,AnglesAlcazar2013}. Alternatively, external mechanisms such as ejections/displacements from the center of the galaxy after gravitational recoils have also been proposed as a mechanism to produce undermassive MBHs \citep{Blecha2011,IzquierdoVillalba2020,DongPaez2025}. In this regard, several observational studies have recently reported candidates of recoiled MBHs \citep[e.g.,][]{vanDokkum2023,Barrows2025,Islam2026}, including what may represent the first confirmed case \citep{vanDokkum2025}. However, the interpretation of some systems remains debated \citep{SanchezAlmeida2023}, and the overall frequency and impact of recoil events are still uncertain.

Within the framework of hierarchical structure formation, all the processes described above are expected to operate concurrently, collectively shaping the full distribution of MBHs around the median $M_{\rm BH}\,{-}\,M_*$ relation and naturally producing both overmassive and undermassive MBH systems. Understanding the emergence of these two populations requires, therefore, tracing how the relative importance of MBH accretion efficiency, galaxy assembly, feedback, and dynamical processes evolves with redshift. In this work, we address this by exploring the emergence of overmassive and undermassive MBHs in the $M_{\rm BH}\,{-}\,M_*$ relation at different galaxy masses and cosmological epochs. To this end, we employ the state-of-the-art semi-analytical model (SAM) \lgbh{}, a flexible and well-tested galaxy formation framework that can be run on top of the merger trees from the \texttt{Millennium} suite of simulations \citep{Bonoli2025}. We stress that our goal is to focus on the physical conditions that give rise to the over- and under-massive systems, irrespective of their observability. We are aware that many low-mass MBHs may fall below current detection thresholds, limiting their identification and preventing a full characterisation of the true scaling relation. As a result, it would be challenging to determine the over- or under-massive nature of MBHs. The present work examines exclusively the physical origin of the overmassive and undermassive MBH regimes, while a dedicated assessment of their detectability and observational biases is deferred to future studies. The paper is organised as follows: In Section~\ref{sec:LgalaxiesBH} we present the merger trees from the \texttt{Millennium} simulation suite, the main features of the \lgbh{} semi-analytical model, and the new prescription implemented for gradual baryonic stripping. In Section~\ref{sec:Results} we present the definition of overmassive and undermassive MBHs and characterise their formation scenarios. Finally, in Section~\ref{sec:Conclusions} we summarize the key findings. A Lambda Cold Dark Matter $(\Lambda$CDM) cosmology with parameters $\Omega_{\rm m} \,{=}\,0.315$, $\Omega_{\rm \Lambda}\,{=}\,0.685$, $\Omega_{\rm b}\,{=}\,0.045$, $\sigma_{8}\,{=}\,0.9$ and $\rm H_0\,{=}\,67.3\, \rm km\,s^{-1}\,Mpc^{-1}$ is adopted throughout the paper \citep{PlanckCollaboration2014}.

%%%%%%%%%%%%%%%%%%%%%%%%%%%%%%%%%%%%%%%%%%%%%%%%%%%%%%%%%%%%%%

\section{The galaxy formation model: \lgbh{} SAM} \label{sec:LgalaxiesBH}
 In this section, we summarise the main physics implemented in the \lgbh{} \citep{Bonoli2025}. \lgbh{} builds on the public \lgalaxies{} SAM \citep{Henriques2015}, which follows the cosmological assembly of galaxies by solving analytical equations along the dark matter (DM) halo merger trees from N-body simulations. Updates to \lgalaxies{} \citep{IzquierdoVillalba2020, IzquierdoVillalba2021, Spinoso2022} have led to \lgbh{}, which allows a detailed treatment of the formation and evolution of single and binary MBHs.
%In this section, we summarise the main physics included in the \lgbh{} semi-analytical model (SAM, \citealt{Bonoli2025}). Briefly, \lgbh{} builds on the public version of \lgalaxies{} SAM \citep{Henriques2015}, which follows the cosmological assembly of galaxies by solving a set of analytical equations along the merger trees of dark matter (DM) halos extracted from N-body simulations. Subsequent updates to the \lgalaxies{} model \citep{IzquierdoVillalba2020, IzquierdoVillalba2021, Spinoso2022} have culminated in the current \lgbh{} framework, which enables a detailed treatment of the formation and evolution of both single and binary MBHs.

\subsection{Dark matter merger trees}

\lgbh{} is designed to work on the merger trees derived from DM only N-body simulations. Here, we use the merger trees extracted from the \texttt{Millennium} (MS, \citealt{Springel2005}) and \texttt{Millennium-II} (MSII, \citealt{Boylan-Kolchin2009}) simulations. The MS follows the evolution of  $2160^3$ DM particles of $8.6\,{\times}\,10^8\,\msun/h$ in a $500\,{\rm Mpc}/h$ box from $z\,{=}\,127$ to the present. MSII tracks the same number of particles but in a $100\,{\rm Mpc}/h$ box with 125 times higher mass resolution ($6.885\,{\times}\,10^6\,\msun/h$). Snapshots were stored at 63 (MS) and 68 (MSII) epochs, with typical time intervals of ${\sim}\,300 \,\rm Myr$ (\lgbh{} will make an internal internal interpolation of ${\sim}5\,{-}\,20$ Myr). Halos and substructures were identified using the friend-of-friend and \texttt{SUBFIND} algorithms and arranged into merger trees with \texttt{L-HALOTREE} \citep{Springel2001}. Both simulations were rescaled to \cite{PlanckCollaboration2014} cosmology following \cite{AnguloandWhite2010} methodology.

Taking into account the above, the MS enables the exploration of large cosmological volumes but at the expense of a limited mass resolution, which limits the study of low-mass galaxies. In contrast, the merger trees of MSII provide sufficient resolution to trace the formation and evolution of small galaxies. However, its smaller volume hinders its capacity to statistically sample rare high-redshift and massive low-redshift systems. To overcome these limitations and exploit the complementary strengths of both simulations, we do not present independent results from MS and MSII. Instead, we combine their merger trees into the ones called \MSG{}, following the grafting procedure described in \citet{Bonoli2025}. This approach proceeds in two steps. First, the merger tree branches of MS are extended below the resolution limit by incorporating higher-resolution information from MSII, thereby recovering merger events that would otherwise be unresolved. Second, when the MS haloes (above and below resolution) are first resolved, they are initialised with representative galaxies drawn from the population of objects generated by \lgbh{} with MSII trees. This ensures that the systems are modelled as already evolved rather than as pristine baryonic reservoirs. Therefore, the \MSG{} enables us to construct merger trees with the same high resolution as the MSII but within the larger volume (and statistics) of the MS. Therefore, unless otherwise stated, all the results presented here will correspond to the ones obtained with the \MSG{} merger trees.

\subsection{Galaxy and MBH formation and evolution} 

This section describes the galaxy and MBH formation framework of \lgbh{} \citep{Henriques2015} and introduces a new prescription for gradual galaxy stripping.
%This section describes the galaxy formation framework implemented in \lgbh{}, which is based on the \lgalaxies{} model of \citet{Henriques2015}. It also introduces a new prescription that accounts for gradual galaxy stripping.

\subsubsection{Formation and assembly of galaxies: Gradual galaxy stripping} \label{sec:galaxy_stripping}

The \lgbh{} model builds on the seminal works of \citet{WhiteandRees1978} and \citet{WhiteFrenk1991}. In this framework, a collapsing DM halo captures a fraction of cosmic baryons, forming a hot gaseous halo. This one can cool down via radiative processes and settle at the halo centre, forming a rotationally supported disc. Continuous cold gas accretion fuels SF, forming a stellar disc component and triggering supernova (SN) feedback, which regulates the galaxy assembly by heating and expelling cold gas. In parallel, gas accretion of the central MBH from the hot atmosphere moderates further gas cooling via the injection of energy. Galactic bulges can form via secular processes, such as disc instabilities (DI) triggered by intense and sustained episodes of SF, or through external events caused by galaxy mergers. Specifically, mergers follow the coalescence of parent DM halos on timescales estimated by \citet{BinneyTremaine2008} and are divided between major and minor: while major mergers ($m_R\,{>}\,0.2$) produce spheroidal remnants, minor mergers ($m_R\,{<}\,0.2$) preserve the primary disc while growing the bulge via satellite accretion. %Finally, to account for the coarse temporal resolution of MS and MSII snapshots, \lgbh{} interpolates between them on timescales of ${\sim}5\,{-}\,20$ Myr.}

 \lgbh{} also accounts for environmental effects, including ram-pressure stripping and tidal disruption, as key mechanisms shaping the galaxy population \citep{Henriques2015,Henriques2020}. Regarding tidal disruption, in the fiducial \lgbh{} implementation, satellite galaxies are assumed to be instantaneously disrupted once tidal forces from the host DM halo become strong enough to unbind them. In this work, we modify this treatment by allowing for the gradual stripping of the satellite stellar and cold-gas components before any complete disruption. Specifically, we assume that the material of the satellite can remain self-bound only out to a radius $R_t$, defined as \citep{King1962,TaylorBabul2001}:
\begin{equation} \label{eq:TidalRadius}
R_t \,{=}\, \left( \frac{G M_{\rm gal}}{\omega^2 - \Phi^{''}_{DM}(r)} \right),
\end{equation}
where $r$ is the instantaneous radial position of the satellite within its host DM halo, $\omega$ is its orbital angular velocity, and $\phi^{''}_{DM}(r)$ is the second radial derivative of the gravitational potential generated by the DM halo, which is assumed to follow a Navarro–Frenk–White (NFW) density profile \citep{NFW1996}. Here, $M_{gal}$ denotes the total baryonic mass of the satellite galaxy (stars plus cold gas). Assuming this $R_t$ definition, any material located beyond this radius is assumed to become unbound from the satellite and subsequently removed from the galaxy. As in the full disruption case, the stripped cold gas is transferred to the hot-gas reservoir of the host halo, while the stripped stellar component is deposited into the diffuse stellar halo. For simplicity, after the stripping process, the remaining cold gas and stellar mass of the galaxy are determined just as the mass enclosed within $R_t$, assuming an exponential profile for discs and a Hernquist profile for the bulge \citep{Freeman1970,Hernquist1990}:

\begin{equation}
    M_{i}^{\rm disc, N} \,{=}\,  M_{i}^{\rm disc} \left[ 1 - \left(1 + \frac{R_t}{R_i} \right) e^{-R_t/R_i} \right], 
\end{equation}

\begin{equation}
 M_{*}^{\rm bulge,N}  =  M_{*}^{\rm bulge}\left[\frac{ \, R_t^{2}}
     {\left( R_t + \frac{ R_b} { 1 + \sqrt{2} } \right)^{2}} \right],
\end{equation}
where $M_i^{\rm disc} \,{=}\,\{ M_{\rm g}^{\rm disc}, M_{*}^{\rm disc}\}$, and $M_{*}^{\rm bulge}$ are the initial masses of the cold gas disc, stellar disc, and stellar bulge, while $M_{i}^{\rm disc, N} \,{=}\, M_{\rm g}^{\rm disc, N} , M_{*}^{\rm disc, N}$, and $M_{*}^{\rm bulge, N}$ are the corresponding new masses after stripping. $R_i\,{=}\,\{R_g, R_*\}$ and $R_b$ are the effective radii of the cold gas, stellar disc, and stellar bulge, respectively. The new effective radii of these components after the stripping ($R_g^N$, $R_*^N$ and $R_b^N$) are derived under the assumption that their central densities are preserved:

\begin{equation}
    R_i^{N} = \left( \frac{M_{i}^{\rm disc,N}}{M^{\rm disc}_{i}} \right)^{1/2} R_i,  \, \, \, \, \, \, i\,{=}\,\{\rm g,*\},
\end{equation}

\begin{equation}
    R_b^{N} = \left( \frac{M_*^{\rm bulge,N}}{M_*^{\rm bulge}} \right)^{1/3} \left( \frac{R_b}{1+\sqrt{2}} \right).
\end{equation}

%\subsection{Massive black holes}

%This section highlights the key physical processes implemented in \lgbh{} for the growth and dynamics of MBHs.

\subsubsection{The formation of the first MBHs}

The MBH seeding framework in \lgbh{} is described in Spinoso et al. (in prep.) and \citet{Bonoli2025}. In summary, it represents an extension of the model proposed by \citet{Spinoso2022}, which introduces multiple formation channels for MBH seeds. The updated scheme allows for the coexisting emergence of four seed types: Population III remnants (PopIII, ${\sim}100\,\msun$), runaway stellar mergers (RSM, ${\sim}10^{3-4}\,\msun$), direct-collapse black holes (DCBHs, ${\sim}10^{5}\,\msun$), and merger-induced DCBHs (miDCBH, ${\sim}8\times10^4\,\msun$). To provide a physical basis for DCBH and RSM seeds, \citet{Spinoso2022} connected their formation with the intergalactic metal enrichment and radiative feedback in the form of Lyman-Werner radiation, both driven by star formation and supernova activity. Given the coarse resolution of the Millennium simulations, Spinoso et al. (in prep.) and \citet{Bonoli2025} implemented a sub-grid probabilistic model for PopIII remnant formation, assigning seeds based on halo mass and redshift, with parameters calibrated to reproduce the MBH occupation statistics of \citet{Spinoso2022}. The initial seed mass of PopIII remnants is inferred from the stellar mass formed during the halo unresolved evolution, assuming a Larson initial mass function (IMF) and a single, short-lived starburst. Between the time of seed formation and the resolution of its host halo, the seed is allowed to grow at the Eddington rate, modelling the unresolved accretion phase. Finally, like RSM and DCBH seeds, the PopIII seeding is terminated once the mean IGM metallicity exceeds a critical value.

\subsubsection{Gas accretion onto MBHs: The emergence of the $M_{\rm BH} \,{-}\,M_*$ relation}

In \lgbh{}, newly formed MBHs are initially assigned a random spin, which is then evolved self-consistently through gas accretion and mergers \citep{IzquierdoVillalba2020}. The growth of MBHs is primarily driven by the accretion of cold gas funneled towards the galactic centre after galaxy mergers or DIs. This inflowing gas accumulates in a central reservoir ($M_{\rm res}$) around the MBH and is accreted in two stages \citep{IzquierdoVillalba2020,IzquierdoVillalba2024}. In the first one, accretion can be Eddington-limited or super-Eddington, depending on the local conditions. Super-Eddington episodes occur when the gas reservoir is large ($\mathcal{R} = \rm M_{res}/M_{BH} \,{>}\, 2 \,{\times}\, 10^4$) and the inflow rate is high ($\rm \dot{M}_{inflow} \,{>}\, 10 \, \msun/yr$). During such events, the  Eddington ratio ($f_{\rm Edd} \,{=}\, \rm L_{bol}/L_{Edd}$) follows the form:
\begin{equation}
f{\rm Edd} = B(\chi) \left[\frac{0.985}{{\rm \dot{M}{Edd}/\dot{M}}+C(\chi)}+\frac{0.015}{{\rm \dot{M}{Edd}/\dot{M}}+D(\chi)}\right],
\end{equation}
where $\dot{M}$ and $\dot{M}_{\rm Edd}$ are the MBH and Eddington accretion rates, and $\rm L{bol}$ and $\rm L_{Edd}$ the corresponding luminosities. The functions $B(\chi)$, $C(\chi)$, and $D(\chi)$ are adopted from \citet{Madau2014}. When these criteria are not met, accretion proceeds at the standard Eddington limit ($f_{\rm Edd} \,{=}\,1$). The second stage begins once the MBH consumes part of the reservoir \citep{Bonoli2025}, with accretion declining as:
\begin{equation}
f_{\rm Edd} = \left[1 + \left((t-t_0)/t_Q\right)^{1/2}\right]^{-2/\beta},
\end{equation}
where $t_Q \,{=}\, t_d \, \xi^\beta / (\beta \ln 10)$ sets the characteristic decline timescale, with $t_d \,{=}\, 1.26 \,{\times}\, 10^8$ yr, $\beta \,{=}\, 0.4$, and $\xi \,{=}\,0.3$, consistent with the self-regulated MBH growth models of \cite{Hopkins2009}. All free parameters are calibrated to reproduce local $M_{\rm BH}-M_*$ scaling relations and the stochastic gravitational wave background measured by PTAs \citep{Antoniadis2023,Agazie2023,Reardon2023}. In this framework, the $M_{\rm BH}-M_*$ relation naturally emerges from the coupled evolution of galaxies and their central MBHs, as accretion is intrinsically linked to processes regulating stellar mass (mergers and DI), producing correlated growth across cosmic time.

\subsubsection{Formation and Evolution of massive black hole binaries}

The \lgbh{} model does not assume that, after a galaxy merger, the satellite MBH instantaneously reaches the centre of the remnant galaxy, forms a bound binary with the central MBH, and coalesces. Instead, it follows the dynamical evolution of the MBH from kiloparsec separations down to parsec scales, where massive black hole binaries (MBHBs) form \citep{IzquierdoVillalba2021}. In brief, after a merger, the satellite MBH is typically deposited several kiloparsecs from the nucleus, where it undergoes a dynamical friction phase \citep{BinneyTremaine2008} that gradually brings it to the center (hereafter referred to as \textit{pairing} MBHs in the \textit{pairing} phase), eventually forming a gravitationally bound MBHB with the central MBH. Once bound, the MBHB orbital separation and eccentricity evolve via coupled equations that account for the surrounding environment \citep{IzquierdoVillalba2021}. In gas-rich systems, the inspiralling and eventual coalescence are driven by circumbinary gas interactions and gravitational wave (GW) emission \citep{Dotti2015,PetersAndMathews1963}, whereas in stellar-dominated systems, three-body interactions with stars (following a Sérsic profile) dominate alongside GWs \citep{Quinlan1997,Sesana2015,PetersAndMathews1963}. In cases of repeated mergers, a third MBH may reach the nucleus before the pre-existing binary coalesces. In such scenarios, \lgbh{} models the resulting triplet dynamics following \citet{Bonetti2018ModelGrid}. The model also accounts for MBH mass growth during both the pairing and hardening phases. Specifically, the satellite MBH can both accumulate gas for accretion during the merger and consume it as it migrates toward the galactic nucleus \citep[see][]{{IzquierdoVillalba2021,IzquierdoVillalba2023,IzquierdoVillalba2023a}}. Once a bound MBHB forms, accretion proceeds according to the preferential accretion approach supported by hydrodynamical simulations \citep[e.g.,][]{DOrazio2013,Farris2014,DOrazio2021}.

\subsubsection{Gravitational wave recoils: Ejections from the galactic center}

%When an MBHB reaches the end of its evolution and the two MBHs merge, \lgbh{} computes the spin of the remnant following \cite{BarausseANDRezzolla2009}, which depends on the mass ratio, spin magnitudes, and relative orientations of the MBH spins and the orbital angular momentum. Since \lgbh{} tracks only the spin magnitude, these orientations remain unconstrained. To address this, the SAM distinguishes between two merger types \citep[see also][]{Barausse2012,Volonteri2013}: \textit{wet} and \textit{dry}. In wet mergers (where the binary mass is smaller than the available gas reservoir), the MBH spins align within approximately $10^\circ$ \citep{Dotti2010}, while in dry mergers they are assumed to be randomly oriented. The magnitude and orientation of the progenitor spins is also important in the model since at coalescence, the anisotropic emission of GWs imparts a recoil to the remnant MBH. \lgbh{} computes this kick velocity using the prescription of \cite{Lousto2012}. If the recoil velocity exceeds the escape speed of the host galaxy, the MBH is expelled and deposited within the dark matter halo, where \lgbh{} subsequently tracks its orbit as presented in \cite{IzquierdoVillalba2020}.

When an MBHB merges, \lgbh{} computes the remnant spin following \cite{BarausseANDRezzolla2009}, which depends on the mass ratio, spin magnitudes, and relative orientations of the MBH spins and the orbital angular momentum. Since \lgbh{} tracks only spin magnitude, orientations remain unconstrained. To overcome this limitation, the SAM distinguishes two merger types \citep[see also][]{Barausse2012,Volonteri2013}: \textit{wet} and \textit{dry}. In wet mergers (where the binary mass is smaller than the available gas reservoir), the MBH spins align within approximately $10^\circ$ \citep{Dotti2010}, while in dry mergers they are assumed to be randomly oriented.  The magnitude and orientation of the progenitor spins is also important to determine the gravitational recoil received by the remnant MBH due to the anisotropic emission of GWs. \lgbh{} computes this kick velocity using the prescription of \cite{Lousto2012}. If the recoil velocity exceeds the escape speed of the host galaxy, the MBH is expelled and deposited within the DM halo, where \lgbh{} subsequently tracks its orbit as presented in \cite{IzquierdoVillalba2020}.

\section{Results} \label{sec:Results}

This section investigates how different physical processes shape the population of overmassive and undermassive MBHs in the $M_{\rm BH}\,{-}\,M_*$ plane within the \lgbh{} model. We stress that our focus is on the physical conditions that give rise to these systems, irrespective of their observability. We are aware that many low-mass MBHs may fall below current detection thresholds, limiting their identification and preventing a full characterisation of the true scaling relation. As a result, it would be challenging to determine the real over- or under-massive nature of MBHs.

To investigate the populations of overmassive and undermassive MBHs at all redshifts, we define five galaxy samples, each corresponding to a specific percentile range of the global $M_{\rm BH}\,{-}\,M_*$ distribution:\\

\noindent \textbf{Median sample}:\\
\noindent (i) \textbf{\textcolor{darkorange}{$1\sigma$ sample}} (hereafter, \Onep{}): Galaxies lying within the central $68\%$ of the global $M_{\rm BH}\,{-}\,M_*$ distribution ($16^{\rm th}$–$84^{\rm th}$ percentiles).\\
\noindent \textbf{Overmassive samples}:\\
\noindent (ii) \textbf{\textcolor{salmon}{$+2\sigma$ sample}} (hereafter, \Twop{}): Galaxies in the upper intermediate tail ($84^{\rm th}$–$97^{\rm th}$ percentiles).\\
\noindent (iii) \textbf{\textcolor{firebrick}{$+3\sigma$ sample}} (hereafter, \Threep{}): Galaxies in the extreme upper tail ($97^{\rm th}$–$100^{\rm th}$ percentiles).\\
\noindent \textbf{Undermassive samples}:\\
\noindent (iv) \textbf{\textcolor{cornflowerblue}{$-2\sigma$ sample}} (hereafter, \Twom{}): Galaxies in the lower intermediate tail ($3^{\rm rd}$–$16^{\rm th}$ percentiles).\\
\noindent (v) \textbf{\textcolor{slateblue}{$-3\sigma$ sample}} (hereafter, \Threem{}): Galaxies in the extreme lower tail ($0^{\rm th}$–$3^{\rm rd}$ percentiles).\\

\begin{table}[]  \label{Table:Summary_Processes}
\small
\centering
\small
\resizebox{\columnwidth}{!}{
\begin{tabular}{llll}

\multicolumn{1}{l}{PROCESS} & STELLAR MASS          & REDSHIFT  & PRODUCTION       \\ \hline \hline
Gradual stripping                                   & All masses              &  $z<2$                     & Overmassive                      \\ \hline
Super-Eddington                                     & $M_*\,{>}\,10^{10} \msun$                     & $z>4$                  & Overmassive                      \\ \hline
                                             & $M_*\,{>}\,10^{9} \msun$                     & All redshifts                     & Undermassive                                             \\ 
\multirow{-2}{*}{GW recoils/Ejections}                        & $M_*\,{<}\,10^{9} \msun$                     &  $z<3$                 & Undermassive                                             \\ \hline
Quiet merger history                                & \multicolumn{1}{l}{$M_*\,{<}\,10^{9} \msun$} & \multicolumn{1}{l}{All redshifts} & \multicolumn{1}{l}{Undermassive} \\ \hline
Quiet secular history                     & \multicolumn{1}{l}{$M_*\,{>}\,10^{10} \msun$} & \multicolumn{1}{l}{All redshifts} & \multicolumn{1}{l}{Undermassive} \\\hline\hline
\end{tabular}
}
\caption{Summary of the physical processes driving the formation of over- and under-massive MBH in the $M_{\rm BH}\,{-}\,M_*$ relation. \textit{Stellar mass} represents the specific stellar mass range where this process is relevant. \textit{Redshift} indicates the redshift range where the process affects the most. Finally, the \textit{Production} indicates which type (over- or under- massive) MBH population produces the process.}
\end{table}

\begin{figure}
    \centering
    \includegraphics[width=1.\columnwidth]{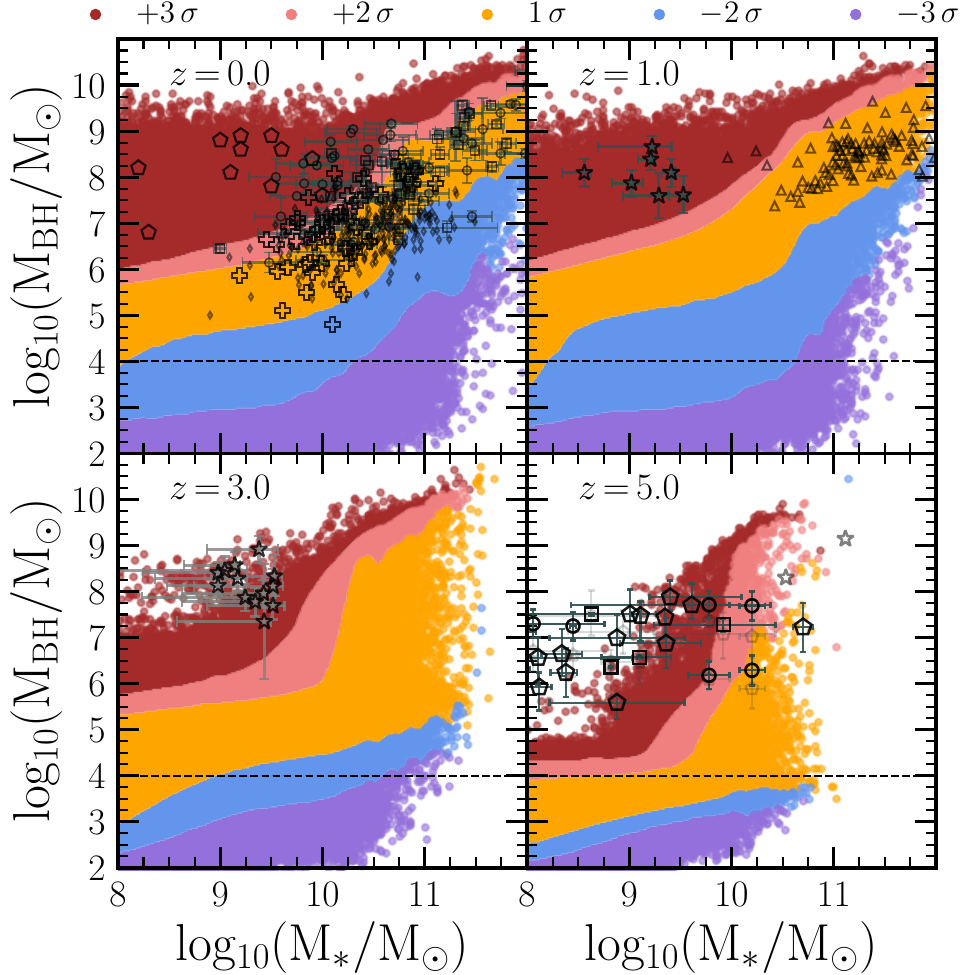}
    \caption{\footnotesize $M_{\rm BH}\,{-}\,M_*$ plane at $z\,{=}\,0,1,3,5$. Different colors represent different populations. The results have been compared with the $z\,{\sim}\,0$ sample of {\protect \cite{Erwin2012}} (squares) {\protect \cite{ReinesVolonteri2015}} (diamonds) and {\protect \cite{Capuzzo2017}} (circles), {\protect \cite{Ferre_Mateu2021}} (pentagons) and {\protect \cite{Ramsden2026} (crosses)}. The $z\,{=}\,1$ and $z\,{=}\,3$ results presents the observational sample of { \protect \cite{Shu2020}} (triangles) and the overmassive MBHs of {\protect \cite{Mezcua2023}} and {\protect \cite{Mezcua2024}}, respectively. The $z\,{=}\,5$ results are compared with {\protect \cite{Maiolino2023}} (circles), {\protect \cite{Harikane2023}} (squares), {\protect \cite{Ding2023}} (stars) and {\protect \cite{Lupi2024}} (pentagons). At this redshift the black (grey) points correspond to $z\,{\geq}\,5$ ($z\,{\leq}\,5$). The horizontal line highlights $\,{\leq}\,10^4\, \msun$, a mass range where electromagnetic information about the central MBH of the galaxy is expected to be undetectable by current observational facilities. }
    \label{fig:Scatter_Evolution}
\end{figure}

The distinction described above enables a direct comparison between galaxies hosting over- and under- massive MBHs, providing a standardized framework to investigate the physical processes driving deviations from the average $M_{\rm BH}\,{-}\,M_*$ relation \footnote{We have verified that the conclusions and trends reported in this paper are robust with respect to sample selection, as similar trends are observed when samples are defined around the median relation using fixed MBH mass offsets of ${\pm}\, \rm 1 \, dex$ (\Twop{} and \Twom{}) and ${\pm}\, \rm 2 \, dex$ (\Threep{} and \Threem{}).}. The main results are presented in the following sections, with a summary of their impact on the formation of over- and under- massive populations given in Table~\ref{Table:Summary_Processes}. In this context, Fig.~\ref{fig:Scatter_Evolution} illustrates the distribution of the five samples in the $M_{\rm BH} \,{-}\,M_*$ plane at $z \,{=}\, 0, 1, 3,$ and $5$\footnote{To guide the reader, the \Threep{} (\Threem) sample accounts for ${\sim}\,20\%$ of the whole overmassive (undermassive) population with $M_*>10^8\, \msun$, independently of redshift. This number drops down to $2\,{-}\,4\%$ if we select only systems with $M_*>10^{10}\, \msun$.}. Overall, the model shows good agreement with the $z \,{\sim}\, 0$ observations of \cite{Capuzzo2017}, \cite{Erwin2012} and \cite{ReinesVolonteri2015} as well as with the measurements of ultra-compact dwarf galaxies presented by \citet{Ferre_Mateu2021} and the MBHs population unveiled via TDE events by \cite{Ramsden2026}. At $z \,{=}\, 1\,{-}\,3$, it also remains consistent with the sample of \cite{Shu2020} and is capable of generating a population of very massive MBHs in low-mass galaxies, as detected by \cite{Mezcua2023,Mezcua2024}. At $z \,{\sim}\, 5$, the model also shows reasonable agreement with the massive galaxy population unveiled by JWST observations. It is noteworthy that, compared to the model version presented in \citet{Bonoli2025}, the inclusion of gradual stripping leads to a depletion of $z\,{\sim}\,5$ MBHs with $M_{\rm BH} \,{>}\, 10^6\,\msun$ in galaxies of $M_* \,{\sim}\, 10^{8}\,\msun$. This hinders the production of a large population of overmassive MBH in low-mass galaxies, as recently observed by JWST. We have verified that this difficulty arises because gas stripping systematically reduces the masses of satellite galaxies at the time of merger, relative to a model without stripping. Consequently, the mass ratios in galaxy mergers are smaller, leading to less efficient gas inflows. Therefore, fewer low-mass galaxies experience super-Eddington accretion events. Indeed, \citet{Bonoli2025} showed that such events in \lgbh{} are responsible for generating the JWST-detected MBHs in low-mass galaxies. Our results therefore indicate that, if current JWST observations are confirmed, the implementation of gradual stripping in our model must be less efficient to allow the formation of overmassive MBHs in high-$z$ low-mass galaxies.

\begin{figure}
    \centering
    \includegraphics[width=1.\columnwidth]{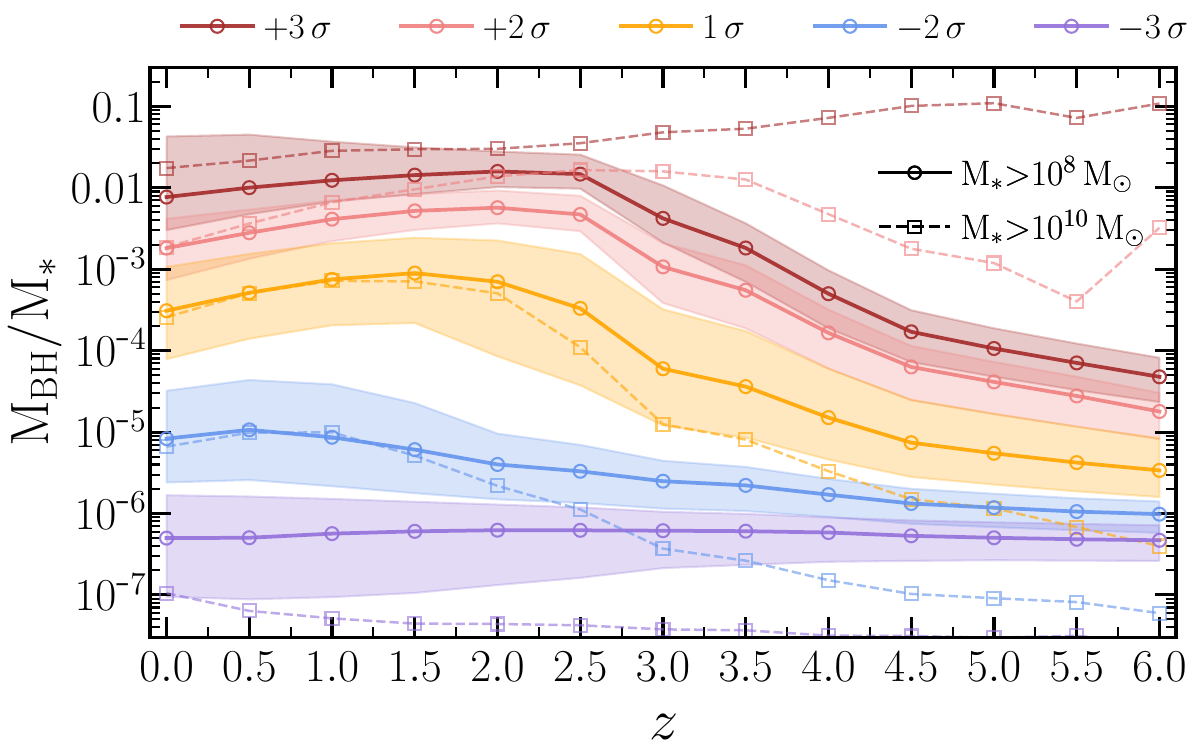}
    \caption{\footnotesize Redshift evolution of the median $M_{\rm BH}/M_*$ ratio for the 5 different samples: \Threep{} (red), \Twop{} (coral),  \Onep{} (orange), \Twom{} (blue) and \Threem{} (purple). Solid lines with circles represent the results for galaxies with $M_*\,{>}\,10^8\, \msun$ while dashed lines with squares represent the same but for galaxies with $M_*\,{>}\,10^{10} \, \msun$. The shaded areas correspond to the $16^{\rm th}\,{-}\,84^{\rm th}$ percentiles.}
    \label{fig:Scaling_Scatter_Evolution}
\end{figure}

To further characterise the behavior of our samples, Fig.~\ref{fig:Scaling_Scatter_Evolution} presents the redshift evolution of the $M_{\rm BH}/M_*$ ratio restricted to galaxies with $M_*\,{>}\,10^{8}\,\msun$. The \Onep{} sample displays a pronounced rise in $M_{\rm BH}/M_*$ from $z \,{\sim}\, 6$ ($M_{\rm BH}/M_* \,{\sim}\, 5\,{\times}\,10^{-6}$) to $z \,{\sim}\, 2$ ($M_{\rm BH}/M_* \,{\sim}\, 10^{-3}$), followed by a decline toward the present epoch, reaching $M_{\rm BH}/M_* \,{\sim}\, 2\,{\times}\,10^{-4}$ at $z\,{=}\,0$. The \Twop{} and \Threep{} samples exhibit similar evolutionary trends, although their $M_{\rm BH}/M_*$ ratios are systematically higher by factors of ${\sim}\,5$–$7$ relative to the \Onep{} population. In contrast, the \Twom{} and \Threem{} samples show little to no evolution: the \Threem{} population maintains a nearly constant ratio of ${\sim}\,7\,{\times}\,10^{-6}$, while the \Twom{} sample remains flat at ${\sim}\,10^{-6}$ up to $z \,{\sim}\, 2$, followed by an increase of about one order of magnitude at lower redshifts. To assess whether these trends persist across different stellar masses, Fig.~\ref{fig:Scaling_Scatter_Evolution} also shows the results for galaxies with $M_* \,{>}\, 10^{10}\,\msun$. The general behaviour of the \Onep{}, \Twom{}, and \Threem{} samples remains broadly consistent with that seen in the lower-mass sample, although their $M_{\rm BH}/M_*$ ratios at $z \,{>}\, 3$ are systematically lower. However, the \Twop{} and \Threep{} populations show a markedly stronger evolution: while their ratios at $z \,{<}\, 2$ are comparable to those of less massive galaxies, they increase dramatically at higher redshifts ($z \,{>}\,3$), by up to four orders of magnitude. For instance, at $z \,{\sim}\, 5$ the \Threep{} (\Twop{}) sample reaches $M_{\rm BH}/M_* \,{\sim}\, 0.1$ (${\sim}\,0.01$), consistent with the overmassive MBH population recently uncovered by JWST observations \citep[see e.g][]{Matthee2024,Harikane2023}.

\begin{figure*}
    \centering
    \includegraphics[width=2.0\columnwidth]{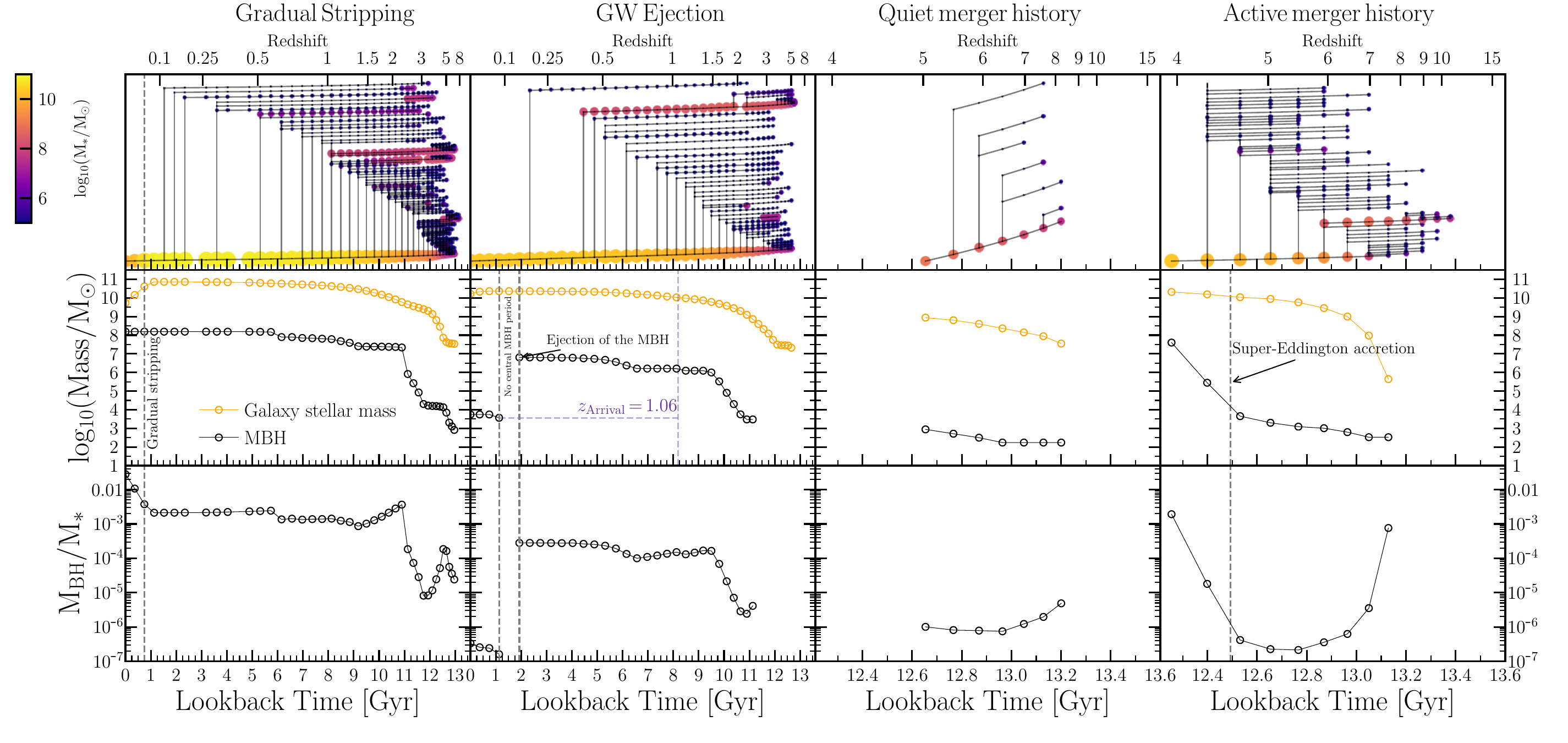}
    \caption{\footnotesize \footnotesize Examples of evolutionary pathways for overmassive and undermassive MBHs. The upper panels show the galaxy merger trees, the middle panels illustrate the assembly of stellar mass and MBH mass, and the lower panels present the evolution of the ratio $M_{\rm BH}/M_*$. The first column corresponds to an overmassive MBH whose growth is driven by the gradual stripping of the host galaxy. The second column shows an undermassive MBH produced by a GW recoil event. The third column depicts an undermassive MBH resulting from a quiescent merger history, while the fourth column presents an overmassive MBH formed through an active merger history combined with a super-Eddington accretion episode.}
    \label{fig:Gal_Tree_Mode}
\end{figure*}

The results shown in Fig.~\ref{fig:Scaling_Scatter_Evolution} indicate that the scaling relation predicted by \lgbh{} exhibits a clear evolution with redshift. At high redshift, the relation also displays a pronounced dependence on galaxy stellar mass. These trends contrast with several previous theoretical studies, which report little to no evolution in the $M_{\rm BH}\,{-}\,M_*$ relation across cosmic time \citep[see][]{Volonteri2016,Huang2018,Habouzit2019,Marshall2020}. Such discrepancies are likely driven by differences in numerical resolution and in the modelling of key physical processes, including MBH growth, MBH seeding, AGN/SN feedback and SF prescriptions \citep[see, e.g.][]{ Habouzit2021}.  A brief comparison of the $M_{\rm BH}/M_*$ ratio between \lgbh{} and five hydrodynamical simulations is provided in Appendix~\ref{Appendix:Comparison_MBH_Mstellar}. However, a detailed investigation of the origin of differences is beyond the scope of this work.\\
In the next sections, we investigate the physical processes that drive deviations from the median $M_{\rm BH}\,{-}\,M_*$ relation, focusing on how overmassive and undermassive MBH populations emerge within the \lgbh{} framework.

\subsection{The role of the environment} \label{sec:Role_Environment}

\begin{figure}
    \centering
    \includegraphics[width=1.0\columnwidth]{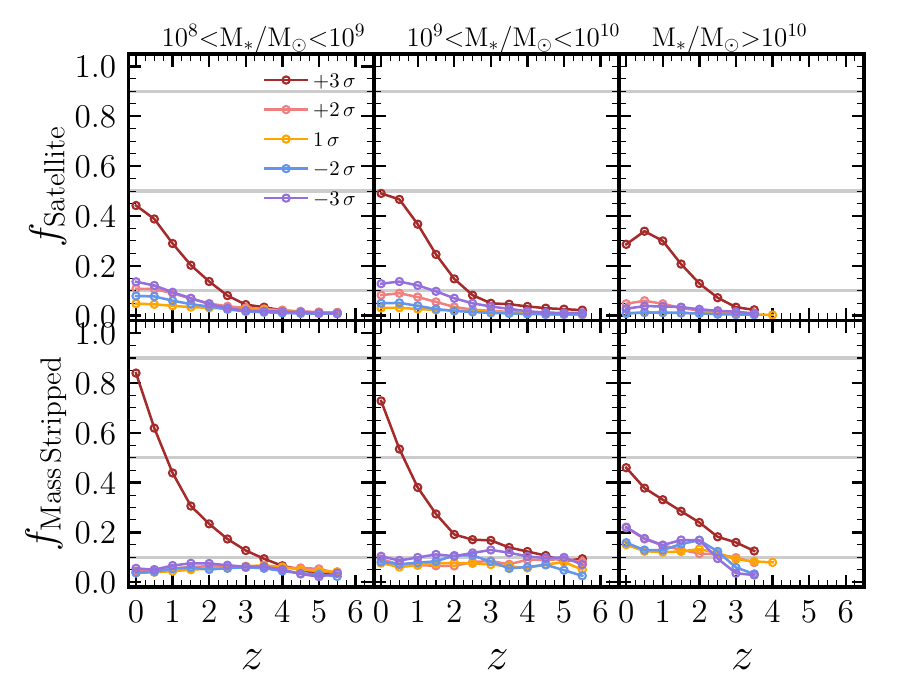}
    \caption{\footnotesize \textbf{Upper panel}: Redshift evolution of the fraction of galaxies that are undergoing a gradual stripping ($f_{\rm Satellite}$). \textbf{Lower panel}: Redshift evolution of the median fraction of stellar mass lost by galaxies due to gradual stripping ($f_{\rm Mass \, Stripped}$). In all panels, there are 5 different samples: \Threep{} (red), \Twop{} (coral),  \Onep{} (orange), \Twom{} (blue) and \Threem{} (purple). The left column corresponds to galaxies with $10^{8} \,{<}\, M_* \,{<}\, 10^{9}\,\msun$, the middle column to those with $10^{9} \,{<}\, M_* \,{<}\, 10^{10}\,\msun$, and the right column to galaxies with $M_* \,{>}\, 10^{10}\,\msun$.}
    \label{fig:StrippingEffect}
\end{figure}

As described in Section~\ref{sec:LgalaxiesBH}, following a halo–halo merger, the galaxy hosted by the smaller halo sinks toward the center of the new host dark matter halo. During this process, it undergoes gradual stripping of stars and cold gas, which can displace the system from its original location in the $M_{\rm BH}\,{-}\,M_*$ relation. This effect is illustrated in the first column of Fig.~\ref{fig:Gal_Tree_Mode}, which shows the evolutionary track of a galaxy and its central MBH extracted from \lgbh{}. As shown, once the galaxy becomes satellite and experiences gradual stripping (indicated by the vertical dashed line), the MBH–galaxy system shifts along the $M_{\rm BH}\,{-}\,M_*$ relation: the MBH mass remains unchanged, while the stellar mass decreases by roughly a factor of eight.

To evaluate the broader role of gradual stripping in generating over- or under-massive MBHs, Fig.~\ref{fig:StrippingEffect} presents the fraction of galaxies in each sample affected by this process. For the \Threem{}, \Twom{}, \Onep{}, and \Twop{} populations, less than 10\% of galaxies experience stellar stripping, independently of redshift or stellar mass. However, the \Threep{} sample exhibits a markedly different behaviour, independently of mass: less than 5\% of galaxies are affected at $z \,{>}\,3$, but this fraction rises steadily toward lower redshifts, from ${\sim}\, 10\%$ at $z \,{\sim}\, 3$ to ${\sim}\, 50\%$ at $z\,{\sim}\,0$. A mild dependence on stellar mass is also evident, with smaller galaxies being slightly more likely to undergo stripping\footnote{Consistent with the higher probability of small galaxies becoming satellites within massive halos.}. Beyond the occurrence of this environmental process, the magnitude of stellar mass loss further differentiates the impact of stripping across samples. This is shown in the lower panel of Fig.~\ref{fig:StrippingEffect}, which reports the median fraction of stellar mass lost by each galaxy since the halo–halo merger. In the \Threem{}, \Twom{}, \Onep{}, and \Twop{} samples, stripping has a negligible effect, with mass losses rarely exceeding 10\%, independent of stellar mass or redshift. Conversely, the \Threep{} sample experiences significantly larger losses, which increase with decreasing redshift and decreasing stellar mass. For galaxies with $10^8 \,{<}\, M_* \,{<}\, 10^9 \, \msun$, the stripped fraction grows from ${\sim}\,20\%$ at $z \,{\sim}\, 3$ to ${\sim}\,90\%$ by $z \,{=}\,0$, with a comparable trend for galaxies in the range $10^9 \,{<}\, M_* \,{<}\, 10^{10}\, \msun$. Galaxies with $M_* \,{>}\, 10^{10}\, \msun$ are less affected, with mass losses reaching at most ${\sim}\, 50\%$ by $z \,{=}\, 0$. We stress that these fractions might remain sensitive to the underlying modelling assumptions about stripping, particularly the predicted galaxy sizes in \lgbh{} and the adopted galaxy mass profiles, which together determine how much mass lies beyond $R_t$ (see Eq.~\eqref{eq:TidalRadius}). Regarding the former, in \lgbh{} galaxy scale lengths are computed self-consistently from angular momentum and energy conservation during galaxy assembly, and are in good agreement with both high-redshift JWST constraints \citep{Herrero-Carrion2026} and low-redshift observations \citep{IzquierdoVillalba2019}, supporting the adopted structural properties. By contrast, the assumed mass profiles (see Section~\ref{sec:galaxy_stripping}) can more directly regulate the stripped fraction in \Threep{} systems: more extended stellar or gas distributions increase the mass beyond $R_t$, enhancing stripping, while more compact profiles have the opposite effect. Despite these dependencies, the qualitative result about gradual stripping leading to the build-up of extreme systems at low redshift is expected to remain robust.\\

Taken together, the results presented in this section indicate that stellar stripping contributes to the scatter in the $M_{\rm BH}\,{-}\,M_*$ relation by producing uniquely (extreme) overmassive MBHs via the reduction of their host galaxies stellar mass (${>}\,20\%$), while leaving the MBH mass unaffected. These results are consistent with the findings of \citet{Weller2023}, who showed that in the \texttt{TNG50} hydrodynamical simulation more than 95\% of satellite galaxies in overdense environments (where tidal stripping is likely to occur) lie above the global $M_{\rm BH}\,{-}\,M_*$ relation. On the observational side, \citet{Ferre_Mateu2021} found that ultracompact dwarfs and compact ellipticals (known to be low-mass systems because of their in situ formation or the stripping of massive galaxies) systematically host overmassive MBHs. Despite these results, our work indicates that environmental process are significant only for ${\sim}\, 40\%$ of the \Threep{} population at $z \,{<}\, 3$ (${\sim}\,10\%$ of the whole overmassive sample), and it cannot account for the overmassive MBHs observed in the \Twop{} population at the same redshift, nor in the \Twop{} and \Threep{} samples at $z \,{>}\, 3$. These findings demonstrate that gradual stellar stripping alone is insufficient to explain the full population of overmassive MBHs, implying that additional physical mechanisms must contribute throughout cosmic time.

\subsection{The role of the ejections after gravitational wave recoils}  \label{sec:Role_Ejections}

\begin{figure}
    \centering
    \includegraphics[width=1.0\columnwidth]{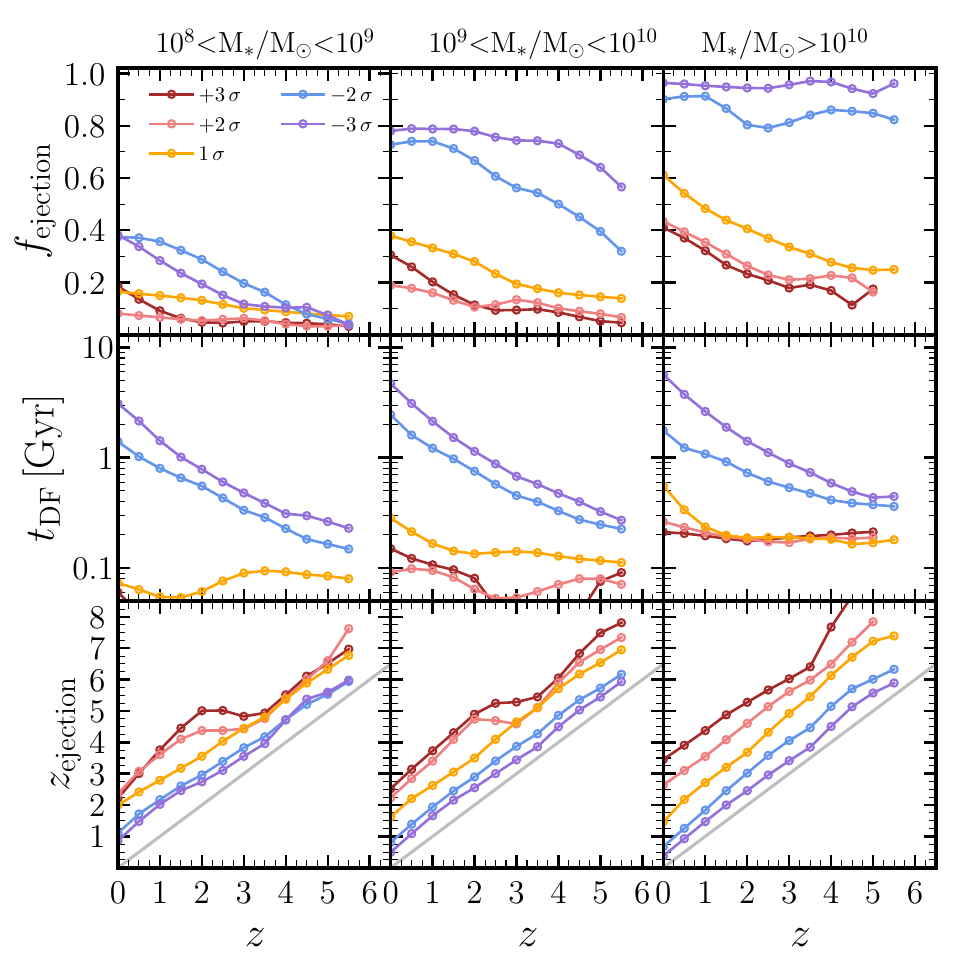}
    \caption{\footnotesize \footnotesize \textbf{First panel}: Redshift evolution of the fraction of galaxies that underwent an ejection of their central MBH due to GW recoils ($f_{\rm ejection}$). \textbf{Second panel}: Median time spent in the dynamical-friction phase ($t_{\rm DF}$) by MBHs that re-filled an empty galactic nucleus due to a GW ejection. \textbf{Third panel}: Redshift evolution of the median redshift at which the MBH was ejected from the center of the galaxy ($z_{\rm ejection}$). The color coding and mass dependence are the same as in Fig.~\ref{fig:StrippingEffect}.
    }
    \label{fig:Recoil_Effect}
\end{figure}

Several theoretical studies have demonstrated that gravitational recoil following MBH coalescences can eject MBHs from their host galaxies, leaving a measurable imprint on galaxy scaling relations \citep{Blecha2011, Gerosa&Sesana2015, IzquierdoVillalba2020, DongPaez2025}. Building on these studies, and to investigate the impact of gravitational recoils and subsequent MBH ejections on the $M_{\rm BH}\,{-}\,M_*$ relation across cosmic time, the upper panel of Fig.~\ref{fig:Recoil_Effect} shows the fraction of galaxies that have experienced at least one central MBH ejection over their lifetime. The results reveal a clear dependence on both stellar mass and sample type. For galaxies with $10^8 \,{<}\, M_* \,{<}\, 10^9\, \msun$, fewer than $20\%$ of systems in the \Onep{}, \Twop{}, and \Threep{} samples undergo an ejection event. The \Twom{} and \Threem{} samples display moderately higher ejection fractions, increasing from ${\sim}\,20\%$ at $z \,{\geq}\, 3$ to ${\sim}\,40\%$ by $z \,{\sim}\, 0$. The relatively low ejection fractions in these $M_* \,{<}\, 10^9 \,\msun$ galaxies are primarily driven by their reduced MBH occupation fraction, which \lgbh{} estimates to be in the range $50\%\,{-}\,80\%$ depending on redshift (see Figure~7 of \citealt{Bonoli2025}). This lowers the probability that galaxy mergers result in MBH coalescence and, therefore, the frequency of producing recoil events. In the intermediate stellar mass range, $10^9 \,{<}\, M_* \,{<}\, 10^{10} \,\msun$, the ejection fractions in the \Onep{}, \Twop{}, and \Threep{} samples show a moderate raise from $10\%$ at $z\,{>}\,2$ to $30\%$ at lower redshifts. In contrast, the \Twom{} and \Threem{} samples exhibit stronger redshift evolution. In the \Twom{} sample, the fraction of galaxies experiencing at least one MBH ejection rises from ${\sim}\,40\%$ at $z \,{\sim}\, 5$ to ${\sim}\,80\%$ at $z \,{<}\, 1$. For the \Threem{} sample, the increase is more rapid, with the ejection fraction already saturating at ${\sim}\,80\%$ by $z\,{\sim}\,4$. Finally, for galaxies with $M_* \,{>}\, 10^{10}\,\msun$, the overall behaviour is qualitatively similar to that observed at lower stellar masses. However, in the \Twom{} and \Threem{} samples, the ejection fractions show little to no redshift dependence, having already reached saturation levels of $90\%\,{-}\,95\%$ by $z \,{\sim}\, 6$. 

The results above indicate that gravitational recoil and the resulting MBH ejection are common processes primarily affecting undermassive MBHs. This has important implications for the evolution of galaxy–MBH systems in the $M_{\rm BH}\,{-}\,M_*$ plane, as the absence of a central MBH after an ejection creates an opportunity for any pairing MBH orbiting within the galaxy (coming from a past galaxy merger) to migrate to the galaxy nucleus and establish itself as the new central MBH\footnote{We have checked that an ejected MBH can subsequently return to the host galaxy and re-establish itself as the central MBH. During the period in which the MBH is displaced, the host galaxy may continue to grow while the MBH stalls, potentially leading to an undermassive MBH upon reincorporation. However, we find that such scenarios are relatively uncommon, occurring in fewer than 10\% of cases. In the majority of systems, the central MBH that refills an empty galactic nucleus after a gravitational ejection is instead an MBH already orbiting within the galaxy, originating from a past galaxy merger.}. This process can reposition the galaxy-MBH system either upward or downward in the $M_{\rm BH}\,{-}\,M_*$ plane, depending on the mass of the incoming MBH. An illustrative example is shown in the second panel of Fig.~\ref{fig:Gal_Tree_Mode}, which shows a galaxy-MBH system in which, following an MBH merger at $z\,{\sim}\,0.15$, the remnant MBH is ejected. After ${\sim}\,1\,\rm Gyr$ of the central MBH ejection, the galaxy nucleus is filled by another MBH that was deposited in the galaxy at $z\,{\sim}\,1$ (i.e ${\sim}\,6\,\rm Gyr$ ago) and is much lighter (${\sim}\,10^4\, \msun$) than expected for the host galaxy stellar mass (${\sim}\,10^{10}\, \msun$). This drastically reduces the galaxy $M_{\rm BH}/M_*$ ratio from ${\sim}\,10^{-4}$ at $z\,{\sim}\,0.25$ to ${\sim}\,10^{-7}$ at $z\,{\sim}\,0$.

The undermassive outcome presented in Fig.~\ref{fig:Gal_Tree_Mode} is not unique to the illustrative example, but instead reflects a general behavior of pairing MBHs in our model. Although \lgbh{} allows pairing MBHs acquired through galaxy mergers to grow, their off-center location causes them to rapidly exhaust their available gas reservoirs, stalls their growth and prevents any sustained co-evolution with their new host galaxies. As a result, when these MBHs eventually settle in a devoid galaxy nucleus, they have masses close to those they had when they were initially acquired during the galaxy merger, typically lower than what would be expected for the new host galaxy. Therefore, the duration of the pairing phase plays a key role in setting the degree of desynchronized growth between the host galaxy and the MBH that eventually repopulates its empty nucleus, and thus determines the final location of the galaxy–MBH system in the $M_{\rm BH}\,{-}\,M_*$ plane (see Appendix~\ref{Appendix:DutyCycle} for the measurement of the galaxy-MBH growth desynchronization via duty cycle analysis).  Motivated by this, it is important to examine how externally acquired MBHs in different samples migrate toward the nucleus in galaxies that temporarily lack a central MBH due to gravitational ejections. This is presented in the second row of Fig.~\ref{fig:Recoil_Effect}, which depicts the time these MBHs spend in the pairing phase within the galaxy before settling at the center. The model predicts that the pairing phases of the \Twom{} and \Threem{} samples varies from $\rm 0.3\, Gyr$ at $z\,{\sim}\,5$ to $3\,\rm Gyr$ at $z\,{\sim}\,1$, independently on the galaxy mass. Conversely, the pairing phase in the \Onep{}, \Twop{}, and \Threep{} samples is typically much shorter, with characteristic durations of ${\sim}\,0.2\,{-}\,0.4\,\rm Gyr$. Interestingly, the duration of the pairing phase tends to be slightly longer in more massive systems, as these galaxies are more extended and MHBs can be deposited at larger galactocentric distances than in lower-mass galaxies. In addition to the duration of the pairing phase, the timing of the gravitational ejection itself constitutes a second critical factor in determining whether a galaxy–MBH system becomes undermassive, overmassive, or converges toward the median relation: even when the pairing phase is short, an ejection occurring close to the galaxy observed redshift may leave insufficient time for the newly established central MBH to grow in synchrony with its host and converge toward the median $M_{\rm BH}\,{-}\,M_*$ relation. To quantify this effect, the third row of Fig.~\ref{fig:Recoil_Effect} shows the redshift of the last MBH ejection for the different samples. In the \Onep{}, \Twop{}, and \Threep{} samples, ejections predominantly occur well before the galaxies reach their observed redshift, with little dependence on stellar mass. Combined with the short pairing phases, this explains why galaxies in the \Onep{}, \Twop{}, and \Threep{} samples can experience gravitational recoils yet remain consistent with the average or overmassive populations: the newly acquired MBHs sink efficiently to the nucleus and have enough time to co-evolve with their hosts before observation. Conversely, Fig.~\ref{fig:Recoil_Effect} indicates that in the \Twom{} and \Threem{} samples, ejections typically occur much closer to the galaxy observed redshift. In these systems, the long pairing phases together with the limited time available for co-evolution inhibit significant convergence toward the median $M_{\rm BH}\,{-}\,M_*$ relation, leaving the systems as undermassive.

\begin{figure}
    \centering
    \includegraphics[width=1.0\columnwidth]{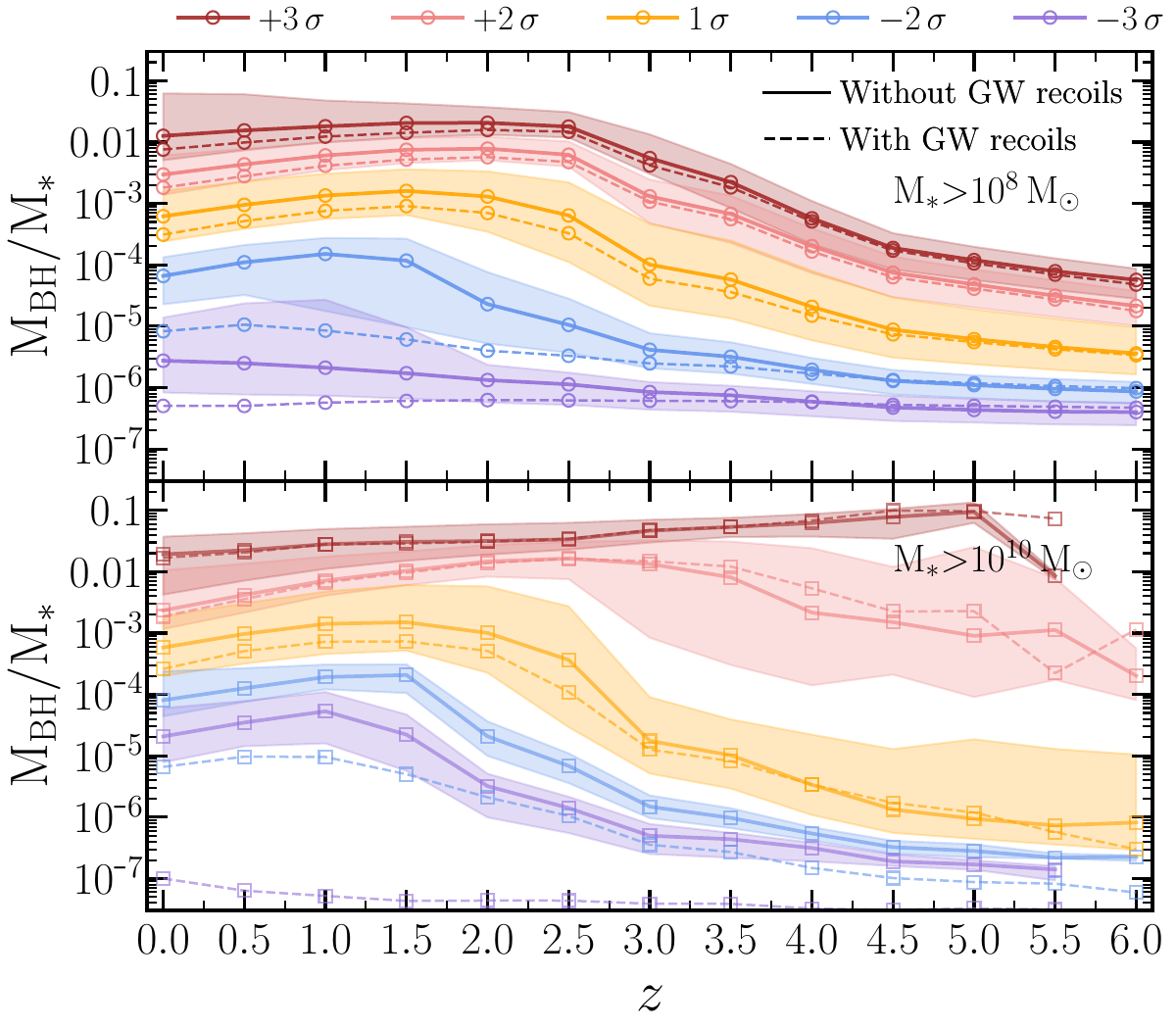}
    \caption{\footnotesize Redshift evolution of the median $M_{\rm BH}/M_*$ ratio for a model in which no GW kicks and ejections are included. The upper (lower) panel corresponds to galaxies with $M_*\,{>}\,10^8\, \msun$ ($M_*\,{>}\,10^{10}\, \msun$) and solid lines correspond to the model without (with) ejections after gravitational recoils. The color coding and mass dependence are the same as in the upper panel of Fig.~\ref{fig:Scaling_Scatter_Evolution}.} 
    \label{fig:Scaling_Scatter_Evolution_NoGWRecoils}
\end{figure}

To link the dynamical processes discussed above with galaxy population statistics, Fig.~\ref{fig:Scaling_Scatter_Evolution_NoGWRecoils} compares the predicted $M_{\rm BH}/M_*$ ratios in models with and without MBH ejections. For galaxies with $M_*\,{>}\, 10^{8}\, \msun$, the absence of gravitational recoils systematically increases the $M_{\rm BH}/M_*$ ratios in the \Onep{}, \Twom{}, and \Threem{} samples, reflecting a reduced fraction of strongly undermassive systems. This effect is particularly pronounced at $z \,{<}\, 3$, where the no-recoil model produces $M_{\rm BH}/M_*$ ratios in the \Twom{} and \Threem{} samples up to 1 dex higher than in the model that includes gravitational ejections. In more massive galaxies ($M_* \,{>}\, 10^{10} \, \msun$), the discrepancy between the two models grows even larger, highlighting that recoil-driven ejections have the strongest impact at the high-mass end of the $M_{\rm BH}\,{-}\,M_*$ relation. 

The analysis presented in this section enables us to conclude that gravitational recoils play a key role in producing exclusively undermassive MBHs in galaxies with $M_* \,{>}\, 10^{10} \,\msun$ at all redshifts, as well as in ${\sim}\,40\%$ of systems with $M_* \,{<}\, 10^{9} \,\msun$ at $z \,{<}\, 3$. The physical mechanism behind the formation of this undermassive population involves ejection events of MBHs after gravitational recoils and the subsequent repopulation of the galactic nucleus by pairing MBHs left over from early galaxy mergers. These newly settled MBHs are usually light as they spend extended periods outside the central regions and only settle into the nucleus close to the observed epoch, leaving them insufficient time to grow in step with their host galaxies. Nonetheless, our results reveal that recoils alone cannot fully explain all undermassive MBHs: even in a no-recoil model, a population of undermassive MBHs persists, particularly among $M_*\,{<}\,10^9\, \msun$ galaxies. In the next section, we analyze how different evolutionary paths contribute to the development of undermassive and the emergence of overmassive MBHs over cosmic time.

\subsection{From a deficit to an excess: Gas Accretion process after mergers and secular evolution} \label{sec:Role_Growth}

\begin{figure}
    \centering
    \includegraphics[width=1.0\columnwidth]{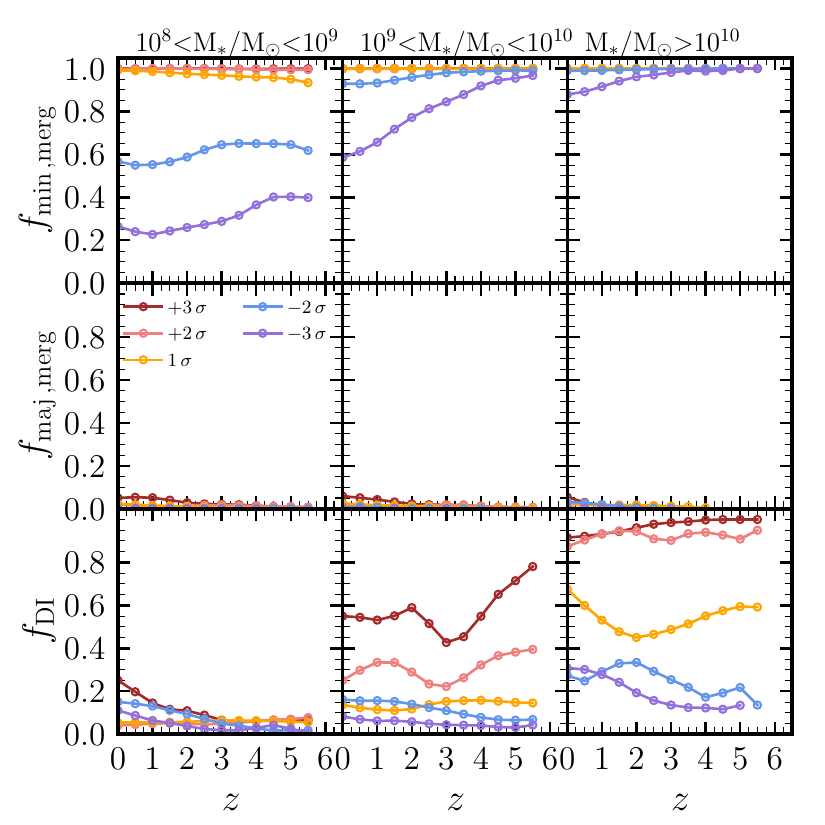}
    \caption{\footnotesize \footnotesize \textbf{Upper panel}: Redshift evolution of the fraction of MBHs in a given stellar mass bin that underwent a growth episode via minor merger ($f_{\rm min\, , merg}$). \textbf{Middle panel}: Redshift evolution of the fraction of MBHs in a given stellar mass bin that experienced a growth phase after a major merger ($f_{\rm maj, \, merg}$). \textbf{Lower panel}: Redshift evolution of the fraction of MBHs in a given stellar mass bin that had a growth phase after a disc instability ($f_{\rm DI}$).  The color coding and mass dependence are the same as in Fig.~\ref{fig:StrippingEffect}. A major merger is defined as a galaxy interaction whose baryonic (gas+stars) mass ratio is larger than 0.2. Otherwise, the interaction is tagged as a minor merger.} 
    \label{fig:MinorMerger_MajorMerger_SecularEvolution}
\end{figure}

Previous sections have demonstrated that stellar mass stripping can account for the emergence of a subset of overmassive MBHs at $z \,{<}\, 3$, while gravitational recoil events explain the presence of undermassive MBHs in galaxies with $M_* \,{>}\, 10^9\, \msun$ across all redshifts. However, these processes alone cannot fully reproduce the overmassive MBH population observed at high redshift, nor the undermassive population in galaxies with $M_* \,{<}\, 10^9 \, \msun$. To investigate the origin of these remaining populations, we now turn to the growth histories of MBHs, which are shaped by both DIs (internal secular processes) and galaxy mergers (external events)\footnote{We have verified that the initial seed mass has a negligible impact on these populations, as nearly all MBHs in the model originate from the light-seed channel associated with Pop~III remnants (see Figure~8 and Figure~7 of \cite{Spinoso2022} and \citep{Bonoli2025}, respectively.}.

\begin{figure}
    \centering
    \includegraphics[width=1.0\columnwidth]{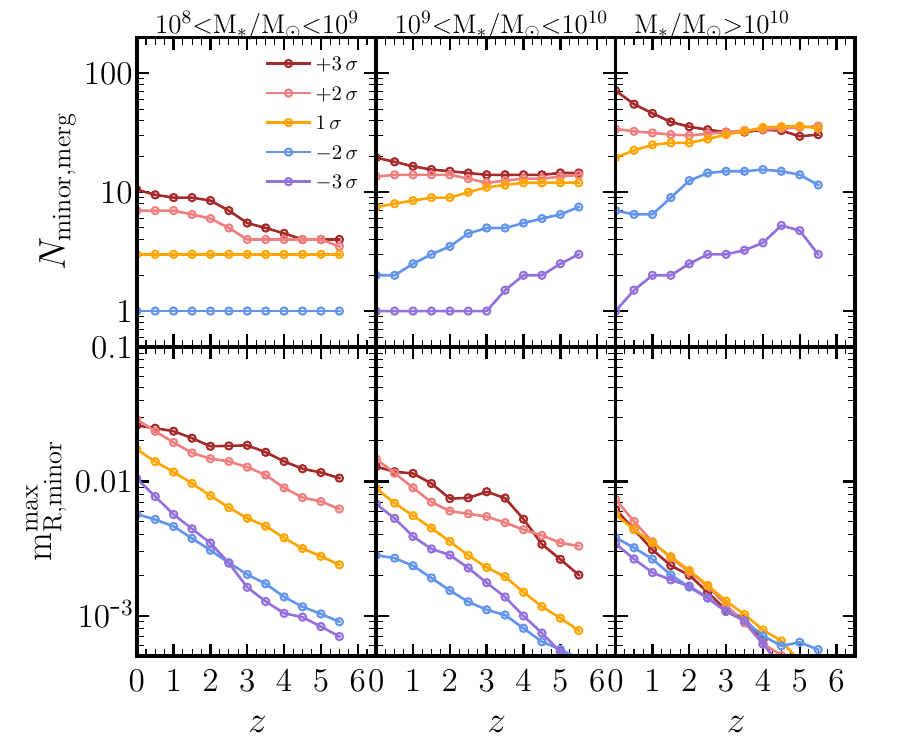}
    \caption{\footnotesize \textbf{Upper panel}: Redshift evolution of the median number of minor mergers that triggered the growth of MBHs hosted in different galaxies ($N_{\rm minor,\, merg}$). \textbf{Lower panel}: Redshift evolution of the median maximum baryonic mass ratio of minor mergers ($\rm m_{R,minor}^{\rm max}$) that triggered the MBH growth in different galaxies.  The color coding and mass dependence are the same as in Fig.~\ref{fig:StrippingEffect}.}
    \label{fig:Number_Minor_Mergers_Mass_Ratio_Mergers}
\end{figure}

\noindent - \textbf{Galaxy mergers}: Focusing on the role of external events, Fig.~\ref{fig:MinorMerger_MajorMerger_SecularEvolution} presents the fraction of MBHs in each sample that have experienced at least one growth episode triggered by major or minor mergers. As shown, major mergers are subdominant across all redshifts and stellar mass bins, with fewer than 5\% of MBHs in any sample undergoing such events. In contrast, minor mergers play a much more significant role. In galaxies with $M_* \,{>}\, 10^9 \,\msun$, over 90\% of MBHs in the \Threep{}, \Twop{}, \Onep{}, and \Twom{} samples, and over 60\% in the \Threem{} sample, have experienced growth episodes triggered by minor mergers, independent of redshift. Despite such a high occurrence rate, the impact of minor mergers on MBHs in these galaxies varies across samples. This is illustrated in Fig.~\ref{fig:Number_Minor_Mergers_Mass_Ratio_Mergers} which shows that MBHs in the \Onep{}, \Twop{}, and \Threep{} samples typically undergo a median of $10\,{-}\,50$ minor merger driven growth episodes, a rate $5\,{-}\,10$ times higher than that of the \Twom{} and \Threem{} populations. This higher frequency of minor merger driven growth is accompanied by systematically larger baryonic mass ratios in the \Threep{}, \Twop{}, and \Onep{} samples, which increase with redshift and saturate at $0.02\,{-}\,0.03$ at $z\,{=}\,0$, roughly a factor of three larger than in the \Threem{} and \Twom{} samples. In the case of MBHs hosted in $M_* \,{<}\, 10^9\, \msun$ galaxies, we observe trends broadly consistent with those found in MBHs hosted in more massive systems, but only for the \Threep{}, \Twop{}, and \Onep{} samples. Conversely, the \Twom{} and \Threem{} populations exhibit significantly fewer minor merger driven growth episodes, with only ${\sim}\,60\%$ and ${\sim} \,20\%$ of cases. Concerning the frequency of MBH growth via minor mergers, the \Threep{} and \Twop{} samples undergo an average of $5\,{-}\,10$ events, whereas the \Twom{} and \Threem{} samples experience typically only $1\,{-}\,3$ events. To illustrate the impact of such limited merger activity on MBH growth, the third row of Fig.~\ref{fig:Gal_Tree_Mode} shows the merger history of a galaxy ($M_*\,{\sim}\,10^9\,\msun$) hosting an undermassive MBH at $z \,{=}\,5$. As seen, the galaxy undergoes few mergers, which restricts the growth of its central MBH, leaving it close to its initial seed mass (${\sim}\,100\,\msun$). Taking together these results imply that minor mergers contribute proportionally more to the growth of the central MBH than to the build-up of the host galaxy stellar mass, thereby having a relevant role in the production of over- and under-massive MBHs.\\   

\begin{figure}
    \centering
    \includegraphics[width=1.0\columnwidth]{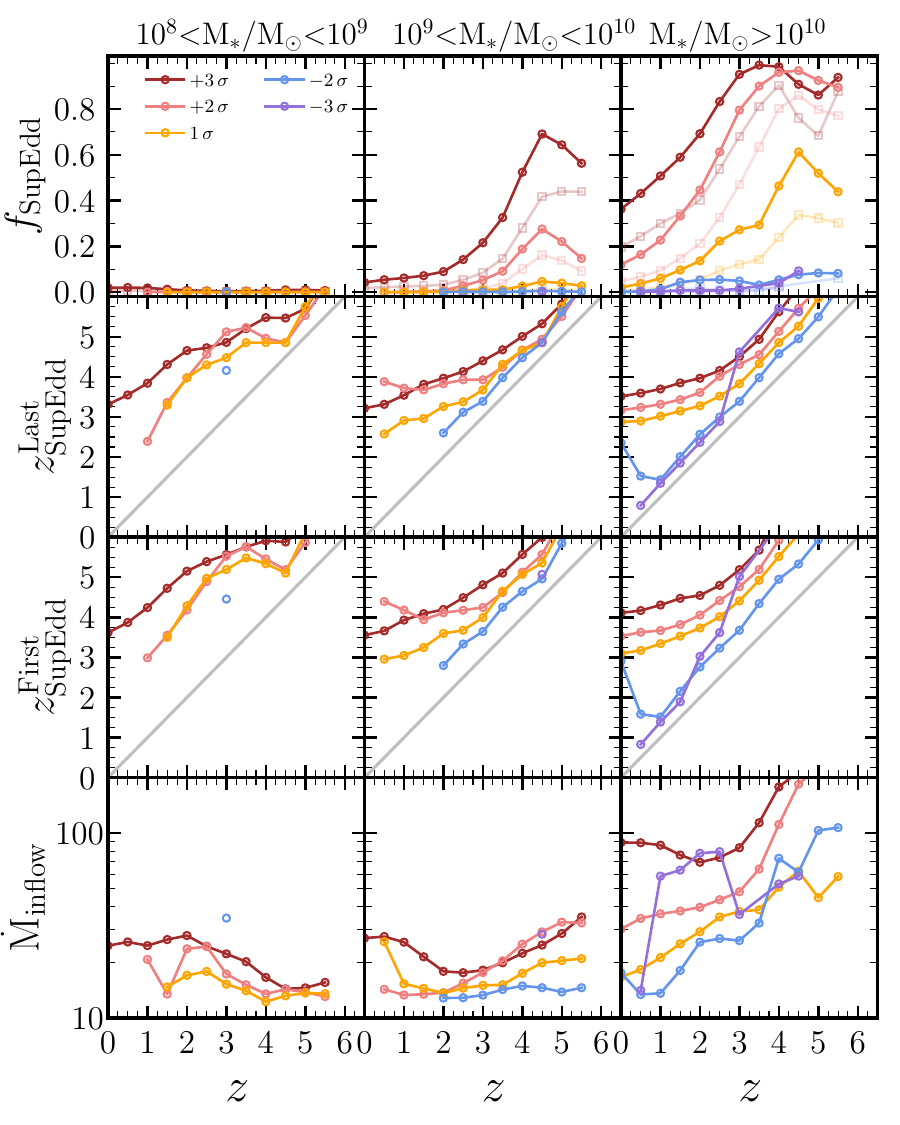}
    \caption{\footnotesize \textbf{First panel}: Redshift evolution of the fraction of galaxies whose central MBH underwent at least one (solid line) or five (pale lines) super-Eddington phases ($f_{\rm SupEdd}$). \textbf{Second panel}: Redshift evolution of the epoch at which MBHs within a given stellar mass bin experienced their most recent super-Eddington accretion phase ($z^{\rm Last}_{\rm SupEdd}$). \textbf{Third panel}: Redshift evolution of the epoch at which MBHs within a given stellar mass bin experienced their first super-Eddington accretion phase ($z^{\rm First}_{\rm SupEdd}$). \textbf{Fourth panel}: Median value of the largest mass inflow ($\rm \dot{M}_{inflow}$) that triggered the super-Eddington phase in the MBH.  The color coding and mass dependence are the same as in Fig.~\ref{fig:StrippingEffect}.} 
    \label{fig:SuperEddington}
\end{figure}

\noindent - \textbf{Disc instabilities}: Beyond the influence of galaxy interactions, secular evolution within galactic discs can trigger DI events, directing gas toward the galactic center and promoting central MBH growth. To quantify the impact of these events in our population of MBHs, the lower row of Fig.~\ref{fig:MinorMerger_MajorMerger_SecularEvolution} shows the fraction of MBHs in different samples that underwent a growth phase driven by DIs. As shown, MBHs residing in $M_* \,{<}\, 10^9 \, \msun$ galaxies receive only a minor contribution from DIs, with fewer than 20\% of MBHs in this mass range growing via such events across all five samples, independently of redshift. For MBHs in galaxies with $M_* \,{>} \, 10^9 \, \msun$, the \Onep{}, \Twom{}, and \Threem{} samples continue to exhibit the same low contribution from DIs. However, DI events play a substantially larger role in the \Twop{} and \Threep{} samples and show a clear dependence on galaxy mass: the fraction of MBHs growing via DIs rises from $30\%\,{-}\,40\%$ in galaxies with $10^9 \,{<}\, M_* \,{<}\, 10^{10} \,\msun$ to nearly 90\% in galaxies with $M_* \,{>}\, 10^{10} \,\msun$, independent of redshift. These results emphasise that, alongside minor mergers, DIs constitute a major channel for MBH growth in the \Twop{} and \Threep{} samples for galaxies with $M_*\,{>}\,10^9\,\msun$ across all redshifts.\\

The trends highlighted above reveal that the specific growth histories of MBHs largely determine the emergence of under- and over-massive systems, accounting for populations that gravitational recoils and stellar stripping alone cannot explain. Undermassive MBHs, primarily comprising the \Twom{} and \Threem{} samples, have experienced exceptionally quiescent evolutionary pathways, characterised by negligible contributions from DIs and only a few galaxy merger events. Such quiet growth histories are most prevalent in low-mass galaxies ($M_* \,{<}\,10^9\, \msun$), where the presence of undermassive MBHs cannot be fully explained by gravitational recoil events alone. In contrast, overmassive MBHs, represented by the \Twop{} and \Threep{} samples, are associated with highly active growth histories, driven by both minor mergers and DI events. This vigorous evolutionary activity becomes increasingly common at higher redshifts, particularly for DI-driven episodes, where stellar stripping alone cannot account for the formation of overmassive MBHs. In fact, in these rapidly evolving environments, efficient gas inflows can drive accretion rates approaching or exceeding the Eddington limit, with super-Eddington phases likely playing a critical role in enabling rapid MBH growth at early cosmic times (see, e.g., spectroscopically identified MBHs in JWST observations, \citealt{Maiolino2023, Harikane2023}). This behaviour is illustrated in the fourth column of Fig.~\ref{fig:Gal_Tree_Mode}, which shows the merger history of a galaxy hosting an overmassive MBH at $z \,{=}\, 4$. As shown, thanks to the rich merger history, the central MBH undergoes a super-Eddington accretion episode, pushing the $M_{\rm BH}/M_*$ ratio from $10^{-6}$ to ${\sim}\,2\,{\times}\,10^{-3}$ in just a few Myrs.

To quantify the role of super-Eddington accretion in shaping overmassive MBHs, Fig.~\ref{fig:SuperEddington} presents the fraction of MBHs experiencing at least one super-Eddington phase. As shown, overmassive MBHs in galaxies with $M_* \,{<}\, 10^9 \, \msun$ rarely undergo such episodes, indicating that the emergence of these systems in low-mass hosts is driven by sustained and frequent growth episodes rather than by short-lived phases of extremely rapid accretion. By contrast, overmassive MBHs in more massive galaxies ($M_* \,{>}\, 10^9 \, \msun$) exhibit a significantly higher incidence of super-Eddington episodes. Specifically, these events are concentrated at $z \,{>}\, 3$, with more than 50\% of MBHs in the \Twop{} and \Threep{} samples experiencing at least one super-Eddington phase. In comparison, the overall MBH population (\Onep{} sample) shows systematically lower fractions, rarely exceeding 30–40\%. Examining the redshift distribution of super-Eddington episodes (second and third panels of Fig.~\ref{fig:SuperEddington}), we can see that MBHs in the \Twop{} and \Threep{} samples begin their rapid growth earlier than those in the \Onep{} sample. Specifically, for overmassive MBHs observed at $z \,{>}\, 3$, both the first and last super-Eddington episodes occur at $z \,{>}\, 5$, whereas for overmassive MBHs observed at $z \,{<}\, 3$, these episodes typically occur around $z \,{\sim}\, 4$. Furthermore, the lower panels of Fig.~\ref{fig:SuperEddington} show that the maximum nuclear inflow rates associated with mergers or DI events triggering super-Eddington phases are consistently higher in the \Twop{} and \Threep{} samples than in the \Onep{} sample. This underscores the importance of super-Eddington accretion in driving the growth of overmassive MBHs. Independent of galaxy mass, high-redshift episodes are more intense, with inflow rates reaching up to twice those observed at lower redshift. Finally, it is worth noticing that the super-Eddington model of \lgbh{} is controlled by different parameters calibrated to reproduce the high-$z$ quasar bolometric luminosity functions \citep{IzquierdoVillalba2024}. Variations that make the criteria for super-Eddington accretion more stringent would reduce the number density of overmassive black holes \citep[see Appendix 1 of][]{IzquierdoVillalba2024}. However, these changes will not alter the qualitative conclusion that high-$z$ overmassive MBHs assemble primarily through active growth episodes in dense, gas-rich environments, where large inflows of fresh gas fuel rapid MBH accretion.

\section{Conclusions} \label{sec:Conclusions}

In this paper, we explore the physical mechanisms driving the formation of overmassive and undermassive MBHs in the $M_{\rm BH}\,{-}\,M_*$ scaling relation across cosmic time and stellar mass. To this end, we have used the state-of-the-art \lgbh{} SAM, an extension of \lgalaxies{} code that self-consistently follows the formation and evolution of single and binary MBHs on top of cosmological merger trees of the \texttt{Millennium} suite of simulations. The model incorporates improved prescriptions for MBH seeding, gas accretion (including super-Eddington phases), stellar and gas stripping, and the full dynamical evolution of MBHs and MBH binaries. The main findings of this work can be summarised as follows:

\begin{itemize}
    \item The model predicts a redshift evolution in the $M_{\rm BH}/M_*$ ratio. For the global population ($M_*\,{>}\,10^8\,\msun$), the value rises from ${\sim}\, 10^{-5}$ at $z\,{>}\,4$ to ${\sim} \, 10^{-3}$ at $z \,{\sim}\, 2$, and then declines to ${\sim}\, 10^{-4}$ by $z \,{=}\,0$. The undermassive population shows little evolution, with ratios remaining between $10^{-6}\,{-}\,10^{-7}$. The overmassive population follows the same trend as the global population, but with ratios up to 1 dex higher. Notably, for this population, there is a mass dependence: high-$z$ galaxies with $M_* \,{>}\, 10^{10}\,\msun$ exhibit larger ratios (${\sim}\, 0.1$) than their $z \,{<}\, 2$ counterparts.\\

    \item  Gradual stripping of the galaxy stellar mass contributes significantly to the emergence of overmassive MBHs at $z \,{<}\, 3$, independently of the galaxy stellar mass. However, this mechanism can explain only ${\sim}\,10\%$ of the low-redshift overmassive population and fails to account for the presence of overmassive MBHs at high redshift.\\

    \item  A rich history of mergers and secular evolution drives frequent and sustained episodes of MBH growth, naturally leading to the formation of overmassive MBHs across all redshifts and stellar masses. In $M_* \,{<}\, 10^9 \,\msun$ galaxies, overmassive MBHs are primarily associated with highly active merger histories, experiencing on average ${\sim}\,10$ interactions (${\sim}\,5$ times more than the average population). In $M_* \,{>}\, 10^9 \,\msun$ galaxies, overmassive MBHs arise from the combination of both vigorous merger activity and frequent disc instabilities driven by secular evolution. Unlike their low-mass counterparts, these massive systems often create the conditions necessary for super-Eddington accretion events, allowing MBHs to grow rapidly and reach overmassive status.\\
        
    \item MBH ejections driven by gravitational recoils play a crucial role in the formation of undermassive MBHs in galaxies with $M_* \,{>}\, 10^9 \, \msun$. In particular, the absence of a central MBH after a gravitational ejection creates an opportunity for any pairing MBH orbiting within the galaxy (coming from a past galaxy merger) to migrate to the galaxy nucleus and establish itself as the new central MBH. The long sinking timescales of these MBHs (often exceeding $\rm 0.3\, Gyr$ at $z\,{\sim}\,5$ and $3\,\rm Gyr$ at $z\,{\sim}\,1$), combined with the fact that gravitational ejections tend to occur shortly before the galaxy is observed, prevent the newly established MBH from growing in step with its host. As a result, these systems remain persistently undermassive relative to the expected $M_{\rm BH}\,{-}\,M_*$ relation.\\

    \item The rise of undermassive MBHs in galaxies with $M_* \,{<}\, 10^9 \,\msun$ cannot be attributed to gravitational-driven MBH ejections, but instead reflects the intrinsically quiescent merger and secular evolutionary histories of their host galaxies. These systems rarely experience disc instability events, and fewer than 60\% undergo galaxy interactions capable of feeding the central MBH. Moreover, both the frequency and the intensity of such mergers are significantly reduced, as these galaxies typically experience only a single interaction, compared to an average of three mergers in the broader population of MBHs hosted in $M_* \,{>}\, 10^9 \,\msun$ galaxies.  
    
\end{itemize}

Our results indicate that the over- and under-massive systems in the $M_{\rm BH}\,{-}\,M_*$ relation arises from the combined, and often competing, effects of: (i) stellar stripping, which shifts systems upward in the $M_{\rm BH}\,{-}\,M_*$ plane (ii) gravitational recoils and MBH ejections, which shift systems downward and (iii) the diversity of MBH fueling histories, where super-Eddington bursts and quiescent growth move MBHs upward and downward, respectively. The interplay of these processes is essential to reproduce both the overmassive MBHs recently observed by JWST and the undermassive MBHs placed in high-mass galaxies. These findings support a picture in which the co-evolution of galaxies and MBHs is fundamentally non-linear and merger-driven, with environmental and dynamical processes playing a key role in shaping the full diversity of MBH populations across cosmic time.

%%%%%%%%%%%%%%%%%%%%%%%%%%%%%%%%%%%%%%%%%%%%%%%%%%%%%%%%%%%%%%
\begin{acknowledgements}
DIV thanks Monica Colpi, Alberto Sessana, Alessandro Lupi, Mar Mezcua, Daniele Spinoso and Silvia Bonoli for their valuable feedback. The project that gave rise to these results received the support of a fellowship from ``la Caixa'' Foundation (ID 100010434). The fellowship code is LCF/BQ/PI25/12100024. DIV also acknowledges the financial support provided under the European Union’s H2020 ERC Consolidator Grant ``Binary Massive Black Hole Astrophysics'' (B Massive, Grant Agreement: 818691) and the European Union Advanced Grant ``PINGU'' (Grant Agreement: 101142079)
\end{acknowledgements}

\bibliographystyle{aa.bst} % style aa.bst
\bibliography{references.bib} % your references Yourfile.bib

@ARTICLE{Madau2014,
       author = {{Madau}, Piero and {Haardt}, Francesco and {Dotti}, Massimo},
        title = "{Super-critical Growth of Massive Black Holes from Stellar-mass Seeds}",
      journal = {\apjl},
         year = 2014,
        month = apr,
       volume = {784},
       number = {2},
          eid = {L38},
        pages = {L38},
          doi = {10.1088/2041-8205/784/2/L38},
archivePrefix = {arXiv},
       eprint = {1402.6995},
 primaryClass = {astro-ph.CO},
       adsurl = {https://ui.adsabs.harvard.edu/abs/2014ApJ...784L..38M}
}

@ARTICLE{Shen2013,
       author = {{Shen}, Yue and {Liu}, Xin and {Loeb}, Abraham and {Tremaine}, Scott},
        title = "{Constraining Sub-parsec Binary Supermassive Black Holes in Quasars with Multi-epoch Spectroscopy. I. The General Quasar Population}",
      journal = {\apj},
         year = 2013,
        month = sep,
       volume = {775},
       number = {1},
          eid = {49},
        pages = {49},
          doi = {10.1088/0004-637X/775/1/49},
archivePrefix = {arXiv},
       eprint = {1306.4330},
 primaryClass = {astro-ph.CO},
       adsurl = {https://ui.adsabs.harvard.edu/abs/2013ApJ...775...49S}
}

@ARTICLE{Hopkins2007,
   author = {{Hopkins}, P.~F. and {Richards}, G.~T. and {Hernquist}, L.},
    title = "{An Observational Determination of the Bolometric Quasar Luminosity Function}",
  journal = {\apj},
   eprint = {astro-ph/0605678},
     year = 2007,
    month = jan,
   volume = 654,
    pages = {731-753},
      doi = {10.1086/509629},
   adsurl = {http://adsabs.harvard.edu/abs/2007ApJ...654..731H}
}

@ARTICLE{Henriques2015,
	author = {{Henriques}, B.~M.~B. and {White}, S.~D.~M. and {Thomas}, P.~A. and 
	{Angulo}, R. and {Guo}, Q. and {Lemson}, G. and {Springel}, V. and 
	{Overzier}, R.},
	title = "{Galaxy formation in the Planck cosmology - I. Matching the observed evolution of star formation rates, colours and stellar masses}",
	journal = {\mnras},
	archivePrefix = "arXiv",
	eprint = {1410.0365},
	year = 2015,
	month = aug,
	volume = 451,
	pages = {2663-2680},
	doi = {10.1093/mnras/stv705},
	adsurl = {http://adsabs.harvard.edu/abs/2015MNRAS.451.2663H}
}

@ARTICLE{MerloniANDHeinz2008,
	author = {{Merloni}, A. and {Heinz}, S.},
	title = "{A synthesis model for AGN evolution: supermassive black holes growth and feedback modes}",
	journal = {\mnras},
	archivePrefix = "arXiv",
	eprint = {0805.2499},
	year = 2008,
	month = aug,
	volume = 388,
	pages = {1011-1030},
	doi = {10.1111/j.1365-2966.2008.13472.x},
	adsurl = {http://adsabs.harvard.edu/abs/2008MNRAS.388.1011M}
}

@ARTICLE{KormendyAndHo2013,
       author = {{Kormendy}, John and {Ho}, Luis C.},
        title = "{Coevolution (Or Not) of Supermassive Black Holes and Host Galaxies}",
      journal = {\araa},
         year = 2013,
        month = aug,
       volume = {51},
       number = {1},
        pages = {511-653},
          doi = {10.1146/annurev-astro-082708-101811},
archivePrefix = {arXiv},
       eprint = {1304.7762},
 primaryClass = {astro-ph.CO},
       adsurl = {https://ui.adsabs.harvard.edu/abs/2013ARA&A..51..511K}
}

@ARTICLE{VestergaardAndPeterson2006,
       author = {{Vestergaard}, Marianne and {Peterson}, Bradley M.},
        title = "{Determining Central Black Hole Masses in Distant Active Galaxies and Quasars. II. Improved Optical and UV Scaling Relationships}",
      journal = {\apj},
         year = 2006,
        month = apr,
       volume = {641},
       number = {2},
        pages = {689-709},
          doi = {10.1086/500572},
archivePrefix = {arXiv},
       eprint = {astro-ph/0601303},
 primaryClass = {astro-ph},
       adsurl = {https://ui.adsabs.harvard.edu/abs/2006ApJ...641..689V}
}

@ARTICLE{Conroy2013,
       author = {{Conroy}, Charlie},
        title = "{Modeling the Panchromatic Spectral Energy Distributions of Galaxies}",
      journal = {\araa},
         year = 2013,
        month = aug,
       volume = {51},
       number = {1},
        pages = {393-455},
          doi = {10.1146/annurev-astro-082812-141017},
archivePrefix = {arXiv},
       eprint = {1301.7095},
 primaryClass = {astro-ph.CO},
       adsurl = {https://ui.adsabs.harvard.edu/abs/2013ARA&A..51..393C}
}

@ARTICLE{Inayoshi2025,
       author = {{Inayoshi}, Kohei},
        title = "{Little Red Dots as the Very First Activity of Black Hole Growth}",
      journal = {\apjl},
         year = 2025,
        month = jul,
       volume = {988},
       number = {1},
          eid = {L22},
        pages = {L22},
          doi = {10.3847/2041-8213/adea66},
archivePrefix = {arXiv},
       eprint = {2503.05537},
 primaryClass = {astro-ph.GA},
       adsurl = {https://ui.adsabs.harvard.edu/abs/2025ApJ...988L..22I}
}

@ARTICLE{McHardy2006,
       author = {{McHardy}, I.~M. and {Koerding}, E. and {Knigge}, C. and {Uttley}, P. and {Fender}, R.~P.},
        title = "{Active galactic nuclei as scaled-up Galactic black holes}",
      journal = {\nat},
         year = 2006,
        month = dec,
       volume = {444},
       number = {7120},
        pages = {730-732},
          doi = {10.1038/nature05389},
archivePrefix = {arXiv},
       eprint = {astro-ph/0612273},
 primaryClass = {astro-ph},
       adsurl = {https://ui.adsabs.harvard.edu/abs/2006Natur.444..730M}
}

@ARTICLE{Mezcua2024,
       author = {{Mezcua}, Mar and {Pacucci}, Fabio and {Suh}, Hyewon and {Siudek}, Malgorzata and {Natarajan}, Priyamvada},
        title = "{Overmassive Black Holes at Cosmic Noon: Linking the Local and the High-redshift Universe}",
      journal = {\apjl},
         year = 2024,
        month = may,
       volume = {966},
       number = {2},
          eid = {L30},
        pages = {L30},
          doi = {10.3847/2041-8213/ad3c2a},
archivePrefix = {arXiv},
       eprint = {2404.05793},
 primaryClass = {astro-ph.GA},
       adsurl = {https://ui.adsabs.harvard.edu/abs/2024ApJ...966L..30M}
}

@ARTICLE{Mezcua2023,
       author = {{Mezcua}, Mar and {Siudek}, Malgorzata and {Suh}, Hyewon and {Valiante}, Rosa and {Spinoso}, Daniele and {Bonoli}, Silvia},
        title = "{Overmassive Black Holes in Dwarf Galaxies Out to z 0.9 in the VIPERS Survey}",
      journal = {\apjl},
         year = 2023,
        month = jan,
       volume = {943},
       number = {1},
          eid = {L5},
        pages = {L5},
          doi = {10.3847/2041-8213/acae25},
archivePrefix = {arXiv},
       eprint = {2212.14057},
 primaryClass = {astro-ph.GA},
       adsurl = {https://ui.adsabs.harvard.edu/abs/2023ApJ...943L...5M}
}

@ARTICLE{VolonteriNatarajan2009,
       author = {{Volonteri}, Marta and {Natarajan}, Priyamvada},
        title = "{Journey to the M$_{BH}$-{\ensuremath{\sigma}} relation: the fate of low-mass black holes in the Universe}",
      journal = {\mnras},
         year = 2009,
        month = dec,
       volume = {400},
       number = {4},
        pages = {1911-1918},
          doi = {10.1111/j.1365-2966.2009.15577.x},
archivePrefix = {arXiv},
       eprint = {0903.2262},
 primaryClass = {astro-ph.CO},
       adsurl = {https://ui.adsabs.harvard.edu/abs/2009MNRAS.400.1911V}
}

@ARTICLE{AnglesAlcazar2013,
       author = {{Angl{\'e}s-Alc{\'a}zar}, Daniel and {{\"O}zel}, Feryal and {Dav{\'e}}, Romeel},
        title = "{Black Hole-Galaxy Correlations without Self-regulation}",
      journal = {\apj},
         year = 2013,
        month = jun,
       volume = {770},
       number = {1},
          eid = {5},
        pages = {5},
          doi = {10.1088/0004-637X/770/1/5},
archivePrefix = {arXiv},
       eprint = {1303.5058},
 primaryClass = {astro-ph.CO},
       adsurl = {https://ui.adsabs.harvard.edu/abs/2013ApJ...770....5A}
}

@ARTICLE{Barausse2012,
	author = {{Barausse}, E.},
	title = "{The evolution of massive black holes and their spins in their galactic hosts}",
	journal = {\mnras},
	archivePrefix = "arXiv",
	eprint = {1201.5888},
	year = 2012,
	month = jul,
	volume = 423,
	pages = {2533-2557},
	doi = {10.1111/j.1365-2966.2012.21057.x},
	adsurl = {http://adsabs.harvard.edu/abs/2012MNRAS.423.2533B}
}

@ARTICLE{BarausseANDRezzolla2009,
	author = {{Barausse}, E. and {Rezzolla}, L.},
	title = "{Predicting the Direction of the Final Spin from the Coalescence of Two Black Holes}",
	journal = {\apjl},
	archivePrefix = "arXiv",
	eprint = {0904.2577},
	primaryClass = "gr-qc",
	year = 2009,
	month = oct,
	volume = 704,
	pages = {L40-L44},
	doi = {10.1088/0004-637X/704/1/L40},
	adsurl = {http://adsabs.harvard.edu/abs/2009ApJ...704L..40B}
}

@ARTICLE{Boylan-Kolchin2009,
   author = {{Boylan-Kolchin}, M. and {Springel}, V. and {White}, S.~D.~M. and 
	{Jenkins}, A. and {Lemson}, G.},
    title = "{Resolving cosmic structure formation with the Millennium-II Simulation}",
  journal = {\mnras},
archivePrefix = "arXiv",
   eprint = {0903.3041},
 primaryClass = "astro-ph.CO",
     year = 2009,
    month = sep,
   volume = 398,
    pages = {1150-1164},
      doi = {10.1111/j.1365-2966.2009.15191.x},
   adsurl = {http://adsabs.harvard.edu/abs/2009MNRAS.398.1150B}
}

@ARTICLE{Volonteri2013,
	author = {{Volonteri}, M. and {Sikora}, M. and {Lasota}, J.-P. and {Merloni}, A.
	},
	title = "{The Evolution of Active Galactic Nuclei and their Spins}",
	journal = {\apj},
	archivePrefix = "arXiv",
	eprint = {1210.1025},
	primaryClass = "astro-ph.HE",
	year = 2013,
	month = oct,
	volume = 775,
	eid = {94},
	pages = {94},
	doi = {10.1088/0004-637X/775/2/94},
	adsurl = {http://adsabs.harvard.edu/abs/2013ApJ...775...94V}
}

@ARTICLE{Springel2001,
	author = {{Springel}, V. and {White}, S.~D.~M. and {Tormen}, G. and {Kauffmann}, G.
	},
	title = "{Populating a cluster of galaxies - I. Results at [formmu2]z=0}",
	journal = {\mnras},
	eprint = {astro-ph/0012055},
	year = 2001,
	month = dec,
	volume = 328,
	pages = {726-750},
	doi = {10.1046/j.1365-8711.2001.04912.x},
	adsurl = {http://adsabs.harvard.edu/abs/2001MNRAS.328..726S}
}

@ARTICLE{Springel2005,
	author = {{Springel}, V.},
	title = "{The cosmological simulation code GADGET-2}",
	journal = {\mnras},
	eprint = {astro-ph/0505010},
	year = 2005,
	month = dec,
	volume = 364,
	pages = {1105-1134},
	doi = {10.1111/j.1365-2966.2005.09655.x},
	adsurl = {http://adsabs.harvard.edu/abs/2005MNRAS.364.1105S}
}

@ARTICLE{AnguloandWhite2010,
   author = {{Angulo}, R.~E. and {White}, S.~D.~M.},
    title = "{One simulation to fit them all - changing the background parameters of a cosmological N-body simulation}",
  journal = {\mnras},
archivePrefix = "arXiv",
   eprint = {0912.4277},
     year = 2010,
    month = jun,
   volume = 405,
    pages = {143-154},
      doi = {10.1111/j.1365-2966.2010.16459.x},
   adsurl = {http://adsabs.harvard.edu/abs/2010MNRAS.405..143A}
}

@ARTICLE{PlanckCollaboration2014,
   author = {{Planck Collaboration} and {Ade}, P.~A.~R. and {Aghanim}, N. and 
	{Armitage-Caplan}, C. and {Arnaud}, M. and {Ashdown}, M. and 
	{Atrio-Barandela}, F. and {Aumont}, J. and {Baccigalupi}, C. and 
	{Banday}, A.~J. and et al.},
    title = "{Planck 2013 results. XVI. Cosmological parameters}",
  journal = {\aap},
archivePrefix = "arXiv",
   eprint = {1303.5076},
     year = 2014,
    month = nov,
   volume = 571,
      eid = {A16},
    pages = {A16},
      doi = {10.1051/0004-6361/201321591},
   adsurl = {http://adsabs.harvard.edu/abs/2014A%26A...571A..16P}
}

@ARTICLE{WhiteFrenk1991,
	author = {{White}, S.~D.~M. and {Frenk}, C.~S.},
	title = "{Galaxy formation through hierarchical clustering}",
	journal = {\apj},
	year = 1991,
	month = sep,
	volume = 379,
	pages = {52-79},
	doi = {10.1086/170483},
	adsurl = {http://adsabs.harvard.edu/abs/1991ApJ...379...52W}
}

@ARTICLE{Croton2006,
   author = {{Croton}, D.~J.},
    title = "{Evolution in the black hole mass-bulge mass relation: a theoretical perspective}",
  journal = {\mnras},
   eprint = {astro-ph/0512375},
     year = 2006,
    month = jul,
   volume = 369,
    pages = {1808-1812},
      doi = {10.1111/j.1365-2966.2006.10429.x},
   adsurl = {http://adsabs.harvard.edu/abs/2006MNRAS.369.1808C}
}

@ARTICLE{Dotti2010,
   author = {{Dotti}, M. and {Volonteri}, M. and {Perego}, A. and {Colpi}, M. and 
	{Ruszkowski}, M. and {Haardt}, F.},
    title = "{Dual black holes in merger remnants - II. Spin evolution and gravitational recoil}",
  journal = {\mnras},
archivePrefix = "arXiv",
   eprint = {0910.5729},
 primaryClass = "astro-ph.HE",
     year = 2010,
    month = feb,
   volume = 402,
    pages = {682-690},
      doi = {10.1111/j.1365-2966.2009.15922.x},
   adsurl = {http://adsabs.harvard.edu/abs/2010MNRAS.402..682D}
}

@ARTICLE{Lousto2012,
   author = {{Lousto}, C.~O. and {Zlochower}, Y. and {Dotti}, M. and {Volonteri}, M.
	},
    title = "{Gravitational recoil from accretion-aligned black-hole binaries}",
  journal = {\prd},
archivePrefix = "arXiv",
   eprint = {1201.1923},
 primaryClass = "gr-qc",
     year = 2012,
    month = apr,
   volume = 85,
   number = 8,
      eid = {084015},
    pages = {084015},
      doi = {10.1103/PhysRevD.85.084015},
   adsurl = {http://adsabs.harvard.edu/abs/2012PhRvD..85h4015L}
}

@ARTICLE{DiMatteo2005,
   author = {{Di Matteo}, T. and {Springel}, V. and {Hernquist}, L.},
    title = "{Energy input from quasars regulates the growth and activity of black holes and their host galaxies}",
  journal = {\nat},
   eprint = {astro-ph/0502199},
     year = 2005,
    month = feb,
   volume = 433,
    pages = {604-607},
      doi = {10.1038/nature03335},
   adsurl = {http://adsabs.harvard.edu/abs/2005Natur.433..604D}
}

@ARTICLE{WhiteandRees1978,
   author = {{White}, S.~D.~M. and {Rees}, M.~J.},
    title = "{Core condensation in heavy halos - A two-stage theory for galaxy formation and clustering}",
  journal = {\mnras},
     year = 1978,
    month = may,
   volume = 183,
    pages = {341-358},
      doi = {10.1093/mnras/183.3.341},
   adsurl = {http://adsabs.harvard.edu/abs/1978MNRAS.183..341W}
}

@ARTICLE{IzquierdoVillalba2019,
       author = {{Izquierdo-Villalba}, David and {Bonoli}, Silvia and {Spinoso}, Daniele and
         {Rosas-Guevara}, Yetli and {Henriques}, Bruno M.~B. and
         {Hern{\'a}ndez-Monteagudo}, Carlos},
        title = "{The build-up of pseudo-bulges in a hierarchical universe}",
      journal = {\mnras},
         year = "2019",
        month = "Sep",
       volume = {488},
       number = {1},
        pages = {609-632},
          doi = {10.1093/mnras/stz1694},
archivePrefix = {arXiv},
       eprint = {1901.10490},
 primaryClass = {astro-ph.GA},
       adsurl = {https://ui.adsabs.harvard.edu/abs/2019MNRAS.488..609I}
}

@ARTICLE{Hernquist1990,
   author = {{Hernquist}, L.},
    title = "{An analytical model for spherical galaxies and bulges}",
  journal = {\apj},
     year = 1990,
    month = jun,
   volume = 356,
    pages = {359-364},
      doi = {10.1086/168845},
   adsurl = {http://adsabs.harvard.edu/abs/1990ApJ...356..359H}
}

@ARTICLE{NFW1996,
   author = {{Navarro}, J.~F. and {Frenk}, C.~S. and {White}, S.~D.~M.},
    title = "{The Structure of Cold Dark Matter Halos}",
  journal = {\apj},
   eprint = {astro-ph/9508025},
     year = 1996,
    month = may,
   volume = 462,
    pages = {563},
      doi = {10.1086/177173},
   adsurl = {http://adsabs.harvard.edu/abs/1996ApJ...462..563N}
}

@ARTICLE{Bonetti2018ModelGrid,
       author = {{Bonetti}, Matteo and {Haardt}, Francesco and {Sesana}, Alberto and {Barausse}, Enrico},
        title = "{Post-Newtonian evolution of massive black hole triplets in galactic nuclei - II. Survey of the parameter space}",
      journal = {\mnras},
         year = 2018,
        month = jul,
       volume = {477},
       number = {3},
        pages = {3910-3926},
          doi = {10.1093/mnras/sty896},
archivePrefix = {arXiv},
       eprint = {1709.06088},
 primaryClass = {astro-ph.GA},
       adsurl = {https://ui.adsabs.harvard.edu/abs/2018MNRAS.477.3910B}
}

@ARTICLE{Ueda2014,
   author = {{Ueda}, Y. and {Akiyama}, M. and {Hasinger}, G. and {Miyaji}, T. and 
	{Watson}, M.~G.},
    title = "{Toward the Standard Population Synthesis Model of the X-Ray Background: Evolution of X-Ray Luminosity and Absorption Functions of Active Galactic Nuclei Including Compton-thick Populations}",
  journal = {\apj},
archivePrefix = "arXiv",
   eprint = {1402.1836},
     year = 2014,
    month = may,
   volume = 786,
      eid = {104},
    pages = {104},
      doi = {10.1088/0004-637X/786/2/104},
   adsurl = {https://ui.adsabs.harvard.edu/abs/2014ApJ...786..104U}
}

@ARTICLE{Aird2015,
       author = {{Aird}, J. and {Coil}, A.~L. and {Georgakakis}, A. and {Nandra}, K. and
         {Barro}, G. and {P{\'e}rez-Gonz{\'a}lez}, P.~G.},
        title = "{The evolution of the X-ray luminosity functions of unabsorbed and absorbed AGNs out to z̃ 5}",
      journal = {\mnras},
         year = "2015",
        month = "Aug",
       volume = {451},
       number = {2},
        pages = {1892-1927},
          doi = {10.1093/mnras/stv1062},
archivePrefix = {arXiv},
       eprint = {1503.01120},
 primaryClass = {astro-ph.HE},
       adsurl = {https://ui.adsabs.harvard.edu/abs/2015MNRAS.451.1892A}
}

@ARTICLE{Savorgnan2016,
       author = {{Savorgnan}, Giulia A.~D. and {Graham}, Alister W. and {Marconi}, Alessand
        ro and {Sani}, Eleonora},
        title = "{Supermassive Black Holes and Their Host Spheroids. II. The Red and Blue Sequence in the M$_{BH}$-M$_{*,sph}$ Diagram}",
      journal = {\apj},
         year = "2016",
        month = "Jan",
       volume = {817},
       number = {1},
          eid = {21},
        pages = {21},
          doi = {10.3847/0004-637X/817/1/21},
archivePrefix = {arXiv},
       eprint = {1511.07437},
 primaryClass = {astro-ph.GA},
       adsurl = {https://ui.adsabs.harvard.edu/abs/2016ApJ...817...21S}
}

@ARTICLE{IzquierdoVillalba2020,
       author = {{Izquierdo-Villalba}, David and {Bonoli}, Silvia and {Dotti}, Massimo and {Sesana}, Alberto and {Rosas-Guevara}, Yetli and {Spinoso}, Daniele},
        title = "{From galactic nuclei to the halo outskirts: tracing supermassive black holes across cosmic history and environments}",
      journal = {\mnras},
         year = 2020,
        month = jul,
       volume = {495},
       number = {4},
        pages = {4681-4706},
          doi = {10.1093/mnras/staa1399},
archivePrefix = {arXiv},
       eprint = {2001.10548},
 primaryClass = {astro-ph.GA},
       adsurl = {https://ui.adsabs.harvard.edu/abs/2020MNRAS.495.4681I}
}

@ARTICLE{Blecha2011,
       author = {{Blecha}, Laura and {Cox}, Thomas J. and {Loeb}, Abraham and
         {Hernquist}, Lars},
        title = "{Recoiling black holes in merging galaxies: relationship to active galactic nucleus lifetimes, starbursts and the M$_{BH}$-{\ensuremath{\sigma}}$_{*}$ relation}",
      journal = {\mnras},
         year = "2011",
        month = "Apr",
       volume = {412},
       number = {4},
        pages = {2154-2182},
          doi = {10.1111/j.1365-2966.2010.18042.x},
archivePrefix = {arXiv},
       eprint = {1009.4940},
 primaryClass = {astro-ph.CO},
       adsurl = {https://ui.adsabs.harvard.edu/abs/2011MNRAS.412.2154B}
}

@ARTICLE{Gerosa&Sesana2015,
       author = {{Gerosa}, Davide and {Sesana}, Alberto},
        title = "{Missing black holes in brightest cluster galaxies as evidence for the occurrence of superkicks in nature}",
      journal = {\mnras},
         year = "2015",
        month = "Jan",
       volume = {446},
       number = {1},
        pages = {38-55},
          doi = {10.1093/mnras/stu2049},
archivePrefix = {arXiv},
       eprint = {1405.2072},
 primaryClass = {astro-ph.GA},
       adsurl = {https://ui.adsabs.harvard.edu/abs/2015MNRAS.446...38G}
}

@ARTICLE{Schmidt1963,
       author = {{Schmidt}, M.},
        title = "{3C 273 : A Star-Like Object with Large Red-Shift}",
      journal = {Nature},
         year = "1963",
        month = "Mar",
       volume = {197},
       number = {4872},
        pages = {1040},
          doi = {10.1038/1971040a0},
       adsurl = {https://ui.adsabs.harvard.edu/abs/1963Natur.197.1040S}
}

@ARTICLE{Quinlan1997,
       author = {{Quinlan}, Gerald D. and {Hernquist}, Lars},
        title = "{The dynamical evolution of massive black hole binaries {\textemdash} II. Self-consistent N-body integrations}",
      journal = {\na},
         year = "1997",
        month = "Dec",
       volume = {2},
       number = {6},
        pages = {533-554},
          doi = {10.1016/S1384-1076(97)00039-0},
archivePrefix = {arXiv},
       eprint = {astro-ph/9706298},
 primaryClass = {astro-ph},
       adsurl = {https://ui.adsabs.harvard.edu/abs/1997NewA....2..533Q}
}

@ARTICLE{SilkAndRees1998,
       author = {{Silk}, Joseph and {Rees}, Martin J.},
        title = "{Quasars and galaxy formation}",
      journal = {\aap},
         year = 1998,
        month = mar,
       volume = {331},
        pages = {L1-L4},
          doi = {10.48550/arXiv.astro-ph/9801013},
archivePrefix = {arXiv},
       eprint = {astro-ph/9801013},
 primaryClass = {astro-ph},
       adsurl = {https://ui.adsabs.harvard.edu/abs/1998A&A...331L...1S}
}

@ARTICLE{MartinNavarro2018,
       author = {{Mart{\'\i}n-Navarro}, Ignacio and {Brodie}, Jean P. and {Romanowsky}, Aaron J. and {Ruiz-Lara}, Tom{\'a}s and {van de Ven}, Glenn},
        title = "{Black-hole-regulated star formation in massive galaxies}",
      journal = {\nat},
         year = 2018,
        month = jan,
       volume = {553},
       number = {7688},
        pages = {307-309},
          doi = {10.1038/nature24999},
archivePrefix = {arXiv},
       eprint = {1801.00807},
 primaryClass = {astro-ph.GA},
       adsurl = {https://ui.adsabs.harvard.edu/abs/2018Natur.553..307M}
}

@ARTICLE{Wild2010,
       author = {{Wild}, Vivienne and {Heckman}, Timothy and {Charlot}, St{\'e}phane},
        title = "{Timing the starburst-AGN connection}",
      journal = {\mnras},
         year = 2010,
        month = jun,
       volume = {405},
       number = {2},
        pages = {933-947},
          doi = {10.1111/j.1365-2966.2010.16536.x},
archivePrefix = {arXiv},
       eprint = {1002.3156},
 primaryClass = {astro-ph.CO},
       adsurl = {https://ui.adsabs.harvard.edu/abs/2010MNRAS.405..933W}
}

@ARTICLE{DeGraf2017,
       author = {{DeGraf}, C. and {Dekel}, A. and {Gabor}, J. and {Bournaud}, F.},
        title = "{Black hole growth and AGN feedback under clumpy accretion}",
      journal = {\mnras},
         year = 2017,
        month = apr,
       volume = {466},
       number = {2},
        pages = {1462-1476},
          doi = {10.1093/mnras/stw2777},
archivePrefix = {arXiv},
       eprint = {1412.3819},
 primaryClass = {astro-ph.GA},
       adsurl = {https://ui.adsabs.harvard.edu/abs/2017MNRAS.466.1462D}
}

@ARTICLE{DeGraf2015,
       author = {{DeGraf}, C. and {Di Matteo}, T. and {Treu}, T. and {Feng}, Y. and {Woo}, J.-H. and {Park}, D.},
        title = "{Scaling relations between black holes and their host galaxies: comparing theoretical and observational measurements, and the impact of selection effects}",
      journal = {\mnras},
         year = 2015,
        month = nov,
       volume = {454},
       number = {1},
        pages = {913-932},
          doi = {10.1093/mnras/stv2002},
archivePrefix = {arXiv},
       eprint = {1412.4133},
 primaryClass = {astro-ph.GA},
       adsurl = {https://ui.adsabs.harvard.edu/abs/2015MNRAS.454..913D}
}

@ARTICLE{Carraro2020,
       author = {{Carraro}, R. and {Rodighiero}, G. and {Cassata}, P. and {Brusa}, M. and {Shankar}, F. and {Baronchelli}, I. and {Daddi}, E. and {Delvecchio}, I. and {Franceschini}, A. and {Griffiths}, R. and {Gruppioni}, C. and {L{\'o}pez-Navas}, E. and {Mancini}, C. and {Marchesi}, S. and {Negrello}, M. and {Puglisi}, A. and {Sani}, E. and {Suh}, H.},
        title = "{Coevolution of black hole accretion and star formation in galaxies up to z = 3.5}",
      journal = {\aap},
         year = 2020,
        month = oct,
       volume = {642},
          eid = {A65},
        pages = {A65},
          doi = {10.1051/0004-6361/201936649},
archivePrefix = {arXiv},
       eprint = {2007.11002},
 primaryClass = {astro-ph.GA},
       adsurl = {https://ui.adsabs.harvard.edu/abs/2020A&A...642A..65C}
}

@ARTICLE{Ishibashi2012,
       author = {{Ishibashi}, W. and {Fabian}, A.~C.},
        title = "{Active galactic nucleus feedback and triggering of star formation in galaxies}",
      journal = {\mnras},
         year = 2012,
        month = dec,
       volume = {427},
       number = {4},
        pages = {2998-3005},
          doi = {10.1111/j.1365-2966.2012.22074.x},
archivePrefix = {arXiv},
       eprint = {1209.1480},
 primaryClass = {astro-ph.GA},
       adsurl = {https://ui.adsabs.harvard.edu/abs/2012MNRAS.427.2998I}
}

@ARTICLE{Hopkins2009,
       author = {{Hopkins}, Philip F. and {Hernquist}, Lars},
        title = "{Quasars Are Not Light Bulbs: Testing Models of Quasar Lifetimes with the Observed Eddington Ratio Distribution}",
      journal = {\apj},
         year = 2009,
        month = jun,
       volume = {698},
       number = {2},
        pages = {1550-1569},
          doi = {10.1088/0004-637X/698/2/1550},
archivePrefix = {arXiv},
       eprint = {0809.3789},
 primaryClass = {astro-ph},
       adsurl = {https://ui.adsabs.harvard.edu/abs/2009ApJ...698.1550H}
}

@ARTICLE{Habouzit2021,
       author = {{Habouzit}, M{\'e}lanie and {Li}, Yuan and {Somerville}, Rachel S. and {Genel}, Shy and {Pillepich}, Annalisa and {Volonteri}, Marta and {Dav{\'e}}, Romeel and {Rosas-Guevara}, Yetli and {McAlpine}, Stuart and {Peirani}, S{\'e}bastien and {Hernquist}, Lars and {Angl{\'e}s-Alc{\'a}zar}, Daniel and {Reines}, Amy and {Bower}, Richard and {Dubois}, Yohan and {Nelson}, Dylan and {Pichon}, Christophe and {Vogelsberger}, Mark},
        title = "{Supermassive black holes in cosmological simulations I: M$_{BH}$ - M$_{{\ensuremath{\star}}}$ relation and black hole mass function}",
      journal = {\mnras},
         year = 2021,
        month = may,
       volume = {503},
       number = {2},
        pages = {1940-1975},
          doi = {10.1093/mnras/stab496},
archivePrefix = {arXiv},
       eprint = {2006.10094},
 primaryClass = {astro-ph.GA},
       adsurl = {https://ui.adsabs.harvard.edu/abs/2021MNRAS.503.1940H}
}

@ARTICLE{KormendyAndRichstone1995,
	author = {{Kormendy}, John and {Richstone}, Douglas},
	title = "{Inward Bound---The Search For Supermassive Black Holes In Galactic Nuclei}",
	journal = {\araa},
	year = 1995,
	month = jan,
	volume = {33},
	pages = {581},
	doi = {10.1146/annurev.aa.33.090195.003053},
	adsurl = {https://ui.adsabs.harvard.edu/abs/1995ARA&A..33..581K}
}

@ARTICLE{MarconiandHunt2003,
	author = {{Marconi}, Alessandro and {Hunt}, Leslie K.},
	title = "{The Relation between Black Hole Mass, Bulge Mass, and Near-Infrared Luminosity}",
	journal = {\apjl},
	year = 2003,
	month = may,
	volume = {589},
	number = {1},
	pages = {L21-L24},
	doi = {10.1086/375804},
	archivePrefix = {arXiv},
	eprint = {astro-ph/0304274},
	primaryClass = {astro-ph},
	adsurl = {https://ui.adsabs.harvard.edu/abs/2003ApJ...589L..21M}
}

@ARTICLE{HaringANDRix2004,
	author = {{H{\"a}ring}, N. and {Rix}, H.-W.},
	title = "{On the Black Hole Mass-Bulge Mass Relation}",
	journal = {\apjl},
	eprint = {astro-ph/0402376},
	year = 2004,
	month = apr,
	volume = 604,
	pages = {L89-L92},
	doi = {10.1086/383567},
	adsurl = {http://adsabs.harvard.edu/abs/2004ApJ...604L..89H}
}

@ARTICLE{Ferrarese2000,
	author = {{Ferrarese}, Laura and {Merritt}, David},
	title = "{A Fundamental Relation between Supermassive Black Holes and Their Host Galaxies}",
	journal = {\apjl},
	year = 2000,
	month = aug,
	volume = {539},
	number = {1},
	pages = {L9-L12},
	doi = {10.1086/312838},
	archivePrefix = {arXiv},
	eprint = {astro-ph/0006053},
	primaryClass = {astro-ph},
	adsurl = {https://ui.adsabs.harvard.edu/abs/2000ApJ...539L...9F}
}

@ARTICLE{Gebhardt2000,
	author = {{Gebhardt}, Karl and {Bender}, Ralf and {Bower}, Gary and
	{Dressler}, Alan and {Faber}, S.~M. and {Filippenko}, Alexei V. and
	{Green}, Richard and {Grillmair}, Carl and {Ho}, Luis C. and
	{Kormendy}, John and {Lauer}, Tod R. and {Magorrian}, John and
	{Pinkney}, Jason and {Richstone}, Douglas and {Tremaine}, Scott},
	title = "{A Relationship between Nuclear Black Hole Mass and Galaxy Velocity Dispersion}",
	journal = {\apjl},
	year = 2000,
	month = aug,
	volume = {539},
	number = {1},
	pages = {L13-L16},
	doi = {10.1086/312840},
	archivePrefix = {arXiv},
	eprint = {astro-ph/0006289},
	primaryClass = {astro-ph},
	adsurl = {https://ui.adsabs.harvard.edu/abs/2000ApJ...539L..13G}
}

@ARTICLE{TaylorBabul2001,
       author = {{Taylor}, James E. and {Babul}, Arif},
        title = "{The Dynamics of Sinking Satellites around Disk Galaxies: A Poor Man's Alternative to High-Resolution Numerical Simulations}",
      journal = {\apj},
         year = 2001,
        month = oct,
       volume = {559},
       number = {2},
        pages = {716-735},
          doi = {10.1086/322276},
archivePrefix = {arXiv},
       eprint = {astro-ph/0012305},
 primaryClass = {astro-ph},
       adsurl = {https://ui.adsabs.harvard.edu/abs/2001ApJ...559..716T}
}

@ARTICLE{King1962,
       author = {{King}, Ivan},
        title = "{The structure of star clusters. I. an empirical density law}",
      journal = {\aj},
         year = 1962,
        month = oct,
       volume = {67},
        pages = {471},
          doi = {10.1086/108756},
       adsurl = {https://ui.adsabs.harvard.edu/abs/1962AJ.....67..471K}
}

@BOOK{BinneyTremaine2008,
       author = {{Binney}, James and {Tremaine}, Scott},
        title = "{Galactic Dynamics: Second Edition}",
         year = 2008,
       adsurl = {https://ui.adsabs.harvard.edu/abs/2008gady.book.....B}
}

@ARTICLE{Dotti2015,
       author = {{Dotti}, M. and {Merloni}, A. and {Montuori}, C.},
        title = "{Linking the fate of massive black hole binaries to the active galactic nuclei luminosity function}",
      journal = {\mnras},
         year = 2015,
        month = apr,
       volume = {448},
       number = {4},
        pages = {3603-3607},
          doi = {10.1093/mnras/stv291},
archivePrefix = {arXiv},
       eprint = {1502.03101},
 primaryClass = {astro-ph.HE},
       adsurl = {https://ui.adsabs.harvard.edu/abs/2015MNRAS.448.3603D}
}

@ARTICLE{Croton2006a,
       author = {{Croton}, Darren J. and {Springel}, Volker and {White}, Simon D.~M. and {De Lucia}, G. and {Frenk}, C.~S. and {Gao}, L. and {Jenkins}, A. and {Kauffmann}, G. and {Navarro}, J.~F. and {Yoshida}, N.},
        title = "{The many lives of active galactic nuclei: cooling flows, black holes and the luminosities and colours of galaxies}",
      journal = {\mnras},
         year = 2006,
        month = jan,
       volume = {365},
       number = {1},
        pages = {11-28},
          doi = {10.1111/j.1365-2966.2005.09675.x},
archivePrefix = {arXiv},
       eprint = {astro-ph/0508046},
 primaryClass = {astro-ph},
       adsurl = {https://ui.adsabs.harvard.edu/abs/2006MNRAS.365...11C}
}

@ARTICLE{Sesana2015,
       author = {{Sesana}, Alberto and {Khan}, Fazeel Mahmood},
        title = "{Scattering experiments meet N-body - I. A practical recipe for the evolution of massive black hole binaries in stellar environments}",
      journal = {\mnras},
         year = 2015,
        month = nov,
       volume = {454},
       number = {1},
        pages = {L66-L70},
          doi = {10.1093/mnrasl/slv131},
archivePrefix = {arXiv},
       eprint = {1505.02062},
 primaryClass = {astro-ph.GA},
       adsurl = {https://ui.adsabs.harvard.edu/abs/2015MNRAS.454L..66S}
}

@ARTICLE{Farris2014,
       author = {{Farris}, Brian D. and {Duffell}, Paul and {MacFadyen}, Andrew I. and {Haiman}, Zoltan},
        title = "{Binary Black Hole Accretion from a Circumbinary Disk: Gas Dynamics inside the Central Cavity}",
      journal = {\apj},
         year = 2014,
        month = mar,
       volume = {783},
       number = {2},
          eid = {134},
        pages = {134},
          doi = {10.1088/0004-637X/783/2/134},
archivePrefix = {arXiv},
       eprint = {1310.0492},
 primaryClass = {astro-ph.HE},
       adsurl = {https://ui.adsabs.harvard.edu/abs/2014ApJ...783..134F}
}

@ARTICLE{DOrazio2021,
       author = {{D'Orazio}, Daniel J. and {Duffell}, Paul C.},
        title = "{Orbital Evolution of Equal-mass Eccentric Binaries due to a Gas Disk: Eccentric Inspirals and Circular Outspirals}",
      journal = {\apjl},
         year = 2021,
        month = jun,
       volume = {914},
       number = {1},
          eid = {L21},
        pages = {L21},
          doi = {10.3847/2041-8213/ac0621},
archivePrefix = {arXiv},
       eprint = {2103.09251},
 primaryClass = {astro-ph.HE},
       adsurl = {https://ui.adsabs.harvard.edu/abs/2021ApJ...914L..21D}
}

@ARTICLE{DOrazio2013,
       author = {{D'Orazio}, Daniel J. and {Haiman}, Zolt{\'a}n and {MacFadyen}, Andrew},
        title = "{Accretion into the central cavity of a circumbinary disc}",
      journal = {\mnras},
         year = 2013,
        month = dec,
       volume = {436},
       number = {4},
        pages = {2997-3020},
          doi = {10.1093/mnras/stt1787},
archivePrefix = {arXiv},
       eprint = {1210.0536},
 primaryClass = {astro-ph.GA},
       adsurl = {https://ui.adsabs.harvard.edu/abs/2013MNRAS.436.2997D}
}

@ARTICLE{Marshall2020,
       author = {{Marshall}, Madeline A. and {Mutch}, Simon J. and {Qin}, Yuxiang and {Poole}, Gregory B. and {Wyithe}, J. Stuart B.},
        title = "{Dark-ages reionization and galaxy formation simulation - XVIII. The high-redshift evolution of black holes and their host galaxies}",
      journal = {\mnras},
         year = 2020,
        month = may,
       volume = {494},
       number = {2},
        pages = {2747-2759},
          doi = {10.1093/mnras/staa936},
archivePrefix = {arXiv},
       eprint = {1910.08124},
 primaryClass = {astro-ph.GA},
       adsurl = {https://ui.adsabs.harvard.edu/abs/2020MNRAS.494.2747M}
}

@ARTICLE{Kormendy1992,
	author = {{Kormendy}, John and {Richstone}, Douglas},
	title = "{Evidence for a Supermassive Black Hole in NGC 3115}",
	journal = {\apj},
	year = 1992,
	month = jul,
	volume = {393},
	pages = {559},
	doi = {10.1086/171528},
	adsurl = {https://ui.adsabs.harvard.edu/abs/1992ApJ...393..559K}
}

@ARTICLE{Kormendy1988a,
	author = {{Kormendy}, John},
	title = "{Evidence for a Supermassive Black Hole in the Nucleus of M31}",
	journal = {\apj},
	year = 1988,
	month = feb,
	volume = {325},
	pages = {128},
	doi = {10.1086/165988},
	adsurl = {https://ui.adsabs.harvard.edu/abs/1988ApJ...325..128K}
}

@ARTICLE{Genzel1987,
	author = {{Genzel}, R. and {Townes}, C.~H.},
	title = "{Physical conditions, dynamics, and mass distribution in the center of the galaxy.}",
	journal = {\araa},
	year = 1987,
	month = jan,
	volume = {25},
	pages = {377-423},
	doi = {10.1146/annurev.aa.25.090187.002113},
	adsurl = {https://ui.adsabs.harvard.edu/abs/1987ARA&A..25..377G}
}

@ARTICLE{Genzel1994,
	author = {{Genzel}, R. and {Hollenbach}, D. and {Townes}, C.~H.},
	title = "{The nucleus of our Galaxy}",
	journal = {Reports on Progress in Physics},
	year = 1994,
	month = may,
	volume = {57},
	number = {5},
	pages = {417-479},
	doi = {10.1088/0034-4885/57/5/001},
	adsurl = {https://ui.adsabs.harvard.edu/abs/1994RPPh...57..417G}
}

@ARTICLE{Dressler1988,
	author = {{Dressler}, Alan and {Richstone}, Douglas O.},
	title = "{Stellar Dynamics in the Nuclei of M31 and M32: Evidence for Massive Black Holes}",
	journal = {\apj},
	year = 1988,
	month = jan,
	volume = {324},
	pages = {701},
	doi = {10.1086/165930},
	adsurl = {https://ui.adsabs.harvard.edu/abs/1988ApJ...324..701D}
}

@ARTICLE{Peterson2004,
       author = {{Peterson}, B.~M. and {Ferrarese}, L. and {Gilbert}, K.~M. and {Kaspi}, S. and {Malkan}, M.~A. and {Maoz}, D. and {Merritt}, D. and {Netzer}, H. and {Onken}, C.~A. and {Pogge}, R.~W. and {Vestergaard}, M. and {Wandel}, A.},
        title = "{Central Masses and Broad-Line Region Sizes of Active Galactic Nuclei. II. A Homogeneous Analysis of a Large Reverberation-Mapping Database}",
      journal = {\apj},
         year = 2004,
        month = oct,
       volume = {613},
       number = {2},
        pages = {682-699},
          doi = {10.1086/423269},
archivePrefix = {arXiv},
       eprint = {astro-ph/0407299},
 primaryClass = {astro-ph},
       adsurl = {https://ui.adsabs.harvard.edu/abs/2004ApJ...613..682P}
}

@ARTICLE{Vestergaard2006,
       author = {{Vestergaard}, Marianne and {Peterson}, Bradley M.},
        title = "{Determining Central Black Hole Masses in Distant Active Galaxies and Quasars. II. Improved Optical and UV Scaling Relationships}",
      journal = {\apj},
         year = 2006,
        month = apr,
       volume = {641},
       number = {2},
        pages = {689-709},
          doi = {10.1086/500572},
archivePrefix = {arXiv},
       eprint = {astro-ph/0601303},
 primaryClass = {astro-ph},
       adsurl = {https://ui.adsabs.harvard.edu/abs/2006ApJ...641..689V}
}

@ARTICLE{IzquierdoVillalba2021,
       author = {{Izquierdo-Villalba}, David and {Sesana}, Alberto and {Bonoli}, Silvia and {Colpi}, Monica},
        title = "{Massive black hole evolution models confronting the n-Hz amplitude of the stochastic gravitational wave background}",
      journal = {\mnras},
         year = 2022,
        month = jan,
       volume = {509},
       number = {3},
        pages = {3488-3503},
          doi = {10.1093/mnras/stab3239},
archivePrefix = {arXiv},
       eprint = {2108.11671},
 primaryClass = {astro-ph.GA},
       adsurl = {https://ui.adsabs.harvard.edu/abs/2022MNRAS.509.3488I}
}

@article{PetersAndMathews1963,
  title = {Gravitational Radiation from Point Masses in a Keplerian Orbit},
  author = {Peters, P. C. and Mathews, J.},
  journal = {Phys. Rev.},
  volume = {131},
  issue = {1},
  pages = {435--440},
  numpages = {0},
  year = {1963},
  month = {Jul},
  publisher = {American Physical Society},
  doi = {10.1103/PhysRev.131.435},
  url = {https://link.aps.org/doi/10.1103/PhysRev.131.435}
}

@ARTICLE{Spinoso2022,
       author = {{Spinoso}, D. and {Bonoli}, S. and {Valiante}, R. and {Schneider}, R. and {Izquierdo-Villalba}, D.},
        title = "{Multiflavour SMBH seeding and evolution in cosmological environments}",
      journal = {\mnras},
         year = 2023,
        month = jan,
       volume = {518},
       number = {3},
        pages = {4672-4692},
          doi = {10.1093/mnras/stac3169},
archivePrefix = {arXiv},
       eprint = {2203.13846},
 primaryClass = {astro-ph.GA},
       adsurl = {https://ui.adsabs.harvard.edu/abs/2023MNRAS.518.4672S}
}

@article{Capuzzo2017,
    author = {Capuzzo-Dolcetta, R. and Tosta e Melo, I.},
    title = "{On the relation between the mass of Compact Massive Objects and their host galaxies}",
    journal = {Monthly Notices of the Royal Astronomical Society},
    volume = {472},
    number = {4},
    pages = {4013-4023},
    year = {2017},
    month = {09},
    issn = {0035-8711},
    doi = {10.1093/mnras/stx2246},
    url = {https://doi.org/10.1093/mnras/stx2246},
    eprint = {https://academic.oup.com/mnras/article-pdf/472/4/4013/21075542/stx2246.pdf},
}

@ARTICLE{Magorrian1998,
       author = {{Magorrian}, John and {Tremaine}, Scott and {Richstone}, Douglas and {Bender}, Ralf and {Bower}, Gary and {Dressler}, Alan and {Faber}, S.~M. and {Gebhardt}, Karl and {Green}, Richard and {Grillmair}, Carl and {Kormendy}, John and {Lauer}, Tod},
        title = "{The Demography of Massive Dark Objects in Galaxy Centers}",
      journal = {\aj},
         year = 1998,
        month = jun,
       volume = {115},
       number = {6},
        pages = {2285-2305},
          doi = {10.1086/300353},
archivePrefix = {arXiv},
       eprint = {astro-ph/9708072},
 primaryClass = {astro-ph},
       adsurl = {https://ui.adsabs.harvard.edu/abs/1998AJ....115.2285M}
}

@ARTICLE{Graham2001,
       author = {{Graham}, Alister W. and {Erwin}, Peter and {Caon}, N. and {Trujillo}, I.},
        title = "{A Correlation between Galaxy Light Concentration and Supermassive Black Hole Mass}",
      journal = {\apjl},
         year = 2001,
        month = dec,
       volume = {563},
       number = {1},
        pages = {L11-L14},
          doi = {10.1086/338500},
archivePrefix = {arXiv},
       eprint = {astro-ph/0111152},
 primaryClass = {astro-ph},
       adsurl = {https://ui.adsabs.harvard.edu/abs/2001ApJ...563L..11G}
}

@ARTICLE{Lupi2024,
       author = {{Lupi}, Alessandro and {Quadri}, Giada and {Volonteri}, Marta and {Colpi}, Monica and {Regan}, John A.},
        title = "{Sustained super-Eddington accretion in high-redshift quasars}",
      journal = {\aap},
         year = 2024,
        month = jun,
       volume = {686},
          eid = {A256},
        pages = {A256},
          doi = {10.1051/0004-6361/202348788},
archivePrefix = {arXiv},
       eprint = {2312.08422},
 primaryClass = {astro-ph.GA},
       adsurl = {https://ui.adsabs.harvard.edu/abs/2024A&A...686A.256L}
}

@ARTICLE{NelsonTNGDataReleas2019,
       author = {{Nelson}, Dylan and {Springel}, Volker and {Pillepich}, Annalisa and {Rodriguez-Gomez}, Vicente and {Torrey}, Paul and {Genel}, Shy and {Vogelsberger}, Mark and {Pakmor}, Ruediger and {Marinacci}, Federico and {Weinberger}, Rainer and {Kelley}, Luke and {Lovell}, Mark and {Diemer}, Benedikt and {Hernquist}, Lars},
        title = "{The IllustrisTNG simulations: public data release}",
      journal = {Computational Astrophysics and Cosmology},
         year = 2019,
        month = may,
       volume = {6},
       number = {1},
          eid = {2},
        pages = {2},
          doi = {10.1186/s40668-019-0028-x},
archivePrefix = {arXiv},
       eprint = {1812.05609},
 primaryClass = {astro-ph.GA},
       adsurl = {https://ui.adsabs.harvard.edu/abs/2019ComAC...6....2N}
}

@ARTICLE{Henriques2020,
       author = {{Henriques}, Bruno M.~B. and {Yates}, Robert M. and {Fu}, Jian and {Guo}, Qi and {Kauffmann}, Guinevere and {Srisawat}, Chaichalit and {Thomas}, Peter A. and {White}, Simon D.~M.},
        title = "{L-GALAXIES 2020: Spatially resolved cold gas phases, star formation, and chemical enrichment in galactic discs}",
      journal = {\mnras},
         year = 2020,
        month = feb,
       volume = {491},
       number = {4},
        pages = {5795-5814},
          doi = {10.1093/mnras/stz3233},
archivePrefix = {arXiv},
       eprint = {2003.05944},
 primaryClass = {astro-ph.GA},
       adsurl = {https://ui.adsabs.harvard.edu/abs/2020MNRAS.491.5795H}
}

@ARTICLE{IzquierdoVillalba2024,
       author = {{Izquierdo-Villalba}, David and {Sesana}, Alberto and {Colpi}, Monica and {Spinoso}, Daniele and {Bonetti}, Matteo and {Bonoli}, Silvia and {Valiante}, Rosa},
        title = "{Connecting low-redshift LISA massive black hole mergers to the nHz stochastic gravitational wave background}",
      journal = {\aap},
         year = 2024,
        month = jun,
       volume = {686},
          eid = {A183},
        pages = {A183},
          doi = {10.1051/0004-6361/202449293},
archivePrefix = {arXiv},
       eprint = {2401.10983},
 primaryClass = {astro-ph.GA},
       adsurl = {https://ui.adsabs.harvard.edu/abs/2024A&A...686A.183I}
}

@ARTICLE{IzquierdoVillalba2023a,
       author = {{Izquierdo-Villalba}, David and {Sesana}, Alberto and {Colpi}, Monica},
        title = "{Unveiling the hosts of parsec-scale massive black hole binaries: morphology and electromagnetic signatures}",
      journal = {\mnras},
         year = 2023,
        month = feb,
       volume = {519},
       number = {2},
        pages = {2083-2100},
          doi = {10.1093/mnras/stac3677},
archivePrefix = {arXiv},
       eprint = {2207.04064},
 primaryClass = {astro-ph.GA},
       adsurl = {https://ui.adsabs.harvard.edu/abs/2023MNRAS.519.2083I}
}

@ARTICLE{IzquierdoVillalba2023,
       author = {{Izquierdo-Villalba}, David and {Colpi}, Monica and {Volonteri}, Marta and {Spinoso}, Daniele and {Bonoli}, Silvia and {Sesana}, Alberto},
        title = "{Properties and merger signatures of galaxies hosting LISA coalescing massive black hole binaries}",
      journal = {\aap},
         year = 2023,
        month = sep,
       volume = {677},
          eid = {A123},
        pages = {A123},
          doi = {10.1051/0004-6361/202347008},
archivePrefix = {arXiv},
       eprint = {2305.16410},
 primaryClass = {astro-ph.GA},
       adsurl = {https://ui.adsabs.harvard.edu/abs/2023A&A...677A.123I}
}

@ARTICLE{Matthee2024,
       author = {{Matthee}, Jorryt and {Naidu}, Rohan P. and {Brammer}, Gabriel and {Chisholm}, John and {Eilers}, Anna-Christina and {Goulding}, Andy and {Greene}, Jenny and {Kashino}, Daichi and {Labbe}, Ivo and {Lilly}, Simon J. and {Mackenzie}, Ruari and {Oesch}, Pascal A. and {Weibel}, Andrea and {Wuyts}, Stijn and {Xiao}, Mengyuan and {Bordoloi}, Rongmon and {Bouwens}, Rychard and {van Dokkum}, Pieter and {Illingworth}, Garth and {Kramarenko}, Ivan and {Maseda}, Michael V. and {Mason}, Charlotte and {Meyer}, Romain A. and {Nelson}, Erica J. and {Reddy}, Naveen A. and {Shivaei}, Irene and {Simcoe}, Robert A. and {Yue}, Minghao},
        title = "{Little Red Dots: An Abundant Population of Faint Active Galactic Nuclei at z {\ensuremath{\sim}} 5 Revealed by the EIGER and FRESCO JWST Surveys}",
      journal = {\apj},
         year = 2024,
        month = mar,
       volume = {963},
       number = {2},
          eid = {129},
        pages = {129},
          doi = {10.3847/1538-4357/ad2345},
archivePrefix = {arXiv},
       eprint = {2306.05448},
 primaryClass = {astro-ph.GA},
       adsurl = {https://ui.adsabs.harvard.edu/abs/2024ApJ...963..129M}
}

@ARTICLE{Erwin2012,
       author = {{Erwin}, Peter and {Gadotti}, Dimitri Alexei},
        title = "{Do Nuclear Star Clusters and Supermassive Black Holes Follow the Same Host-Galaxy Correlations?}",
      journal = {Advances in Astronomy},
         year = 2012,
        month = jan,
       volume = {2012},
          eid = {946368},
        pages = {946368},
          doi = {10.1155/2012/946368},
archivePrefix = {arXiv},
       eprint = {1112.2740},
 primaryClass = {astro-ph.CO},
       adsurl = {https://ui.adsabs.harvard.edu/abs/2012AdAst2012E...4E}
}

@ARTICLE{ReinesVolonteri2015,
       author = {{Reines}, Amy E. and {Volonteri}, Marta},
        title = "{Relations between Central Black Hole Mass and Total Galaxy Stellar Mass in the Local Universe}",
      journal = {\apj},
         year = 2015,
        month = nov,
       volume = {813},
       number = {2},
          eid = {82},
        pages = {82},
          doi = {10.1088/0004-637X/813/2/82},
archivePrefix = {arXiv},
       eprint = {1508.06274},
 primaryClass = {astro-ph.GA},
       adsurl = {https://ui.adsabs.harvard.edu/abs/2015ApJ...813...82R}
}

@ARTICLE{Maiolino2023,
       author = {{Maiolino}, Roberto and {Scholtz}, Jan and {Curtis-Lake}, Emma and {Carniani}, Stefano and {Baker}, William and {de Graaff}, Anna and {Tacchella}, Sandro and {{\"U}bler}, Hannah and {D'Eugenio}, Francesco and {Witstok}, Joris and {Curti}, Mirko and {Arribas}, Santiago and {Bunker}, Andrew J. and {Charlot}, St{\'e}phane and {Chevallard}, Jacopo and {Eisenstein}, Daniel J. and {Egami}, Eiichi and {Ji}, Zhiyuan and {Jones}, Gareth C. and {Lyu}, Jianwei and {Rawle}, Tim and {Robertson}, Brant and {Rujopakarn}, Wiphu and {Perna}, Michele and {Sun}, Fengwu and {Venturi}, Giacomo and {Williams}, Christina C. and {Willott}, Chris},
        title = "{JADES: The diverse population of infant black holes at 4 < z < 11: Merging, tiny, poor, but mighty}",
      journal = {\aap},
         year = 2024,
        month = nov,
       volume = {691},
          eid = {A145},
        pages = {A145},
          doi = {10.1051/0004-6361/202347640},
archivePrefix = {arXiv},
       eprint = {2308.01230},
 primaryClass = {astro-ph.GA},
       adsurl = {https://ui.adsabs.harvard.edu/abs/2024A&A...691A.145M}
}

@ARTICLE{Harikane2023,
       author = {{Harikane}, Yuichi and {Zhang}, Yechi and {Nakajima}, Kimihiko and {Ouchi}, Masami and {Isobe}, Yuki and {Ono}, Yoshiaki and {Hatano}, Shun and {Xu}, Yi and {Umeda}, Hiroya},
        title = "{A JWST/NIRSpec First Census of Broad-line AGNs at z = 4-7: Detection of 10 Faint AGNs with M $_{BH}$ {}10$^{6}$-{}10$^{8}$ M $_{{\ensuremath{\odot}}}$ and Their Host Galaxy Properties}",
      journal = {\apj},
         year = 2023,
        month = dec,
       volume = {959},
       number = {1},
          eid = {39},
        pages = {39},
          doi = {10.3847/1538-4357/ad029e},
archivePrefix = {arXiv},
       eprint = {2303.11946},
 primaryClass = {astro-ph.GA},
       adsurl = {https://ui.adsabs.harvard.edu/abs/2023ApJ...959...39H}
}

@ARTICLE{Ding2023,
       author = {{Ding}, Xuheng and {Onoue}, Masafusa and {Silverman}, John D. and {Matsuoka}, Yoshiki and {Izumi}, Takuma and {Strauss}, Michael A. and {Jahnke}, Knud and {Phillips}, Camryn L. and {Li}, Junyao and {Volonteri}, Marta and {Haiman}, Zoltan and {Andika}, Irham Taufik and {Aoki}, Kentaro and {Baba}, Shunsuke and {Bieri}, Rebekka and {Bosman}, Sarah E.~I. and {Bottrell}, Connor and {Eilers}, Anna-Christina and {Fujimoto}, Seiji and {Habouzit}, Melanie and {Imanishi}, Masatoshi and {Inayoshi}, Kohei and {Iwasawa}, Kazushi and {Kashikawa}, Nobunari and {Kawaguchi}, Toshihiro and {Kohno}, Kotaro and {Lee}, Chien-Hsiu and {Lupi}, Alessandro and {Lyu}, Jianwei and {Nagao}, Tohru and {Overzier}, Roderik and {Schindler}, Jan-Torge and {Schramm}, Malte and {Shimasaku}, Kazuhiro and {Toba}, Yoshiki and {Trakhtenbrot}, Benny and {Trebitsch}, Maxime and {Treu}, Tommaso and {Umehata}, Hideki and {Venemans}, Bram P. and {Vestergaard}, Marianne and {Walter}, Fabian and {Wang}, Feige and {Yang}, Jinyi},
        title = "{Detection of stellar light from quasar host galaxies at redshifts above 6}",
      journal = {\nat},
         year = 2023,
        month = sep,
       volume = {621},
       number = {7977},
        pages = {51-55},
          doi = {10.1038/s41586-023-06345-5},
archivePrefix = {arXiv},
       eprint = {2211.14329},
 primaryClass = {astro-ph.GA},
       adsurl = {https://ui.adsabs.harvard.edu/abs/2023Natur.621...51D}
}

@ARTICLE{Agazie2023,
       author = {{Agazie}, Gabriella and {Anumarlapudi}, Akash and {Archibald}, Anne M. and {Arzoumanian}, Zaven and {Baker}, Paul T. and {B{\'e}csy}, Bence and {Blecha}, Laura and {Brazier}, Adam and {Brook}, Paul R. and {Burke-Spolaor}, Sarah and {Burnette}, Rand and {Case}, Robin and {Charisi}, Maria and {Chatterjee}, Shami and {Chatziioannou}, Katerina and {Cheeseboro}, Belinda D. and {Chen}, Siyuan and {Cohen}, Tyler and {Cordes}, James M. and {Cornish}, Neil J. and {Crawford}, Fronefield and {Cromartie}, H. Thankful and {Crowter}, Kathryn and {Cutler}, Curt J. and {Decesar}, Megan E. and {Degan}, Dallas and {Demorest}, Paul B. and {Deng}, Heling and {Dolch}, Timothy and {Drachler}, Brendan and {Ellis}, Justin A. and {Ferrara}, Elizabeth C. and {Fiore}, William and {Fonseca}, Emmanuel and {Freedman}, Gabriel E. and {Garver-Daniels}, Nate and {Gentile}, Peter A. and {Gersbach}, Kyle A. and {Glaser}, Joseph and {Good}, Deborah C. and {G{\"u}ltekin}, Kayhan and {Hazboun}, Jeffrey S. and {Hourihane}, Sophie and {Islo}, Kristina and {Jennings}, Ross J. and {Johnson}, Aaron D. and {Jones}, Megan L. and {Kaiser}, Andrew R. and {Kaplan}, David L. and {Kelley}, Luke Zoltan and {Kerr}, Matthew and {Key}, Joey S. and {Klein}, Tonia C. and {Laal}, Nima and {Lam}, Michael T. and {Lamb}, William G. and {Lazio}, T. Joseph W. and {Lewandowska}, Natalia and {Littenberg}, Tyson B. and {Liu}, Tingting and {Lommen}, Andrea and {Lorimer}, Duncan R. and {Luo}, Jing and {Lynch}, Ryan S. and {Ma}, Chung-Pei and {Madison}, Dustin R. and {Mattson}, Margaret A. and {McEwen}, Alexander and {McKee}, James W. and {McLaughlin}, Maura A. and {McMann}, Natasha and {Meyers}, Bradley W. and {Meyers}, Patrick M. and {Mingarelli}, Chiara M.~F. and {Mitridate}, Andrea and {Natarajan}, Priyamvada and {Ng}, Cherry and {Nice}, David J. and {Ocker}, Stella Koch and {Olum}, Ken D. and {Pennucci}, Timothy T. and {Perera}, Benetge B.~P. and {Petrov}, Polina and {Pol}, Nihan S. and {Radovan}, Henri A. and {Ransom}, Scott M. and {Ray}, Paul S. and {Romano}, Joseph D. and {Sardesai}, Shashwat C. and {Schmiedekamp}, Ann and {Schmiedekamp}, Carl and {Schmitz}, Kai and {Schult}, Levi and {Shapiro-Albert}, Brent J. and {Siemens}, Xavier and {Simon}, Joseph and {Siwek}, Magdalena S. and {Stairs}, Ingrid H. and {Stinebring}, Daniel R. and {Stovall}, Kevin and {Sun}, Jerry P. and {Susobhanan}, Abhimanyu and {Swiggum}, Joseph K. and {Taylor}, Jacob and {Taylor}, Stephen R. and {Turner}, Jacob E. and {Unal}, Caner and {Vallisneri}, Michele and {van Haasteren}, Rutger and {Vigeland}, Sarah J. and {Wahl}, Haley M. and {Wang}, Qiaohong and {Witt}, Caitlin A. and {Young}, Olivia and {Nanograv Collaboration}},
        title = "{The NANOGrav 15 yr Data Set: Evidence for a Gravitational-wave Background}",
      journal = {\apjl},
         year = 2023,
        month = jul,
       volume = {951},
       number = {1},
          eid = {L8},
        pages = {L8},
          doi = {10.3847/2041-8213/acdac6},
archivePrefix = {arXiv},
       eprint = {2306.16213},
 primaryClass = {astro-ph.HE},
       adsurl = {https://ui.adsabs.harvard.edu/abs/2023ApJ...951L...8A}
}

@ARTICLE{Antoniadis2023,
       author = {{EPTA Collaboration} and {InPTA Collaboration} and {Antoniadis}, J. and {Arumugam}, P. and {Arumugam}, S. and {Babak}, S. and {Bagchi}, M. and {Bak Nielsen}, A. -S. and {Bassa}, C.~G. and {Bathula}, A. and {Berthereau}, A. and {Bonetti}, M. and {Bortolas}, E. and {Brook}, P.~R. and {Burgay}, M. and {Caballero}, R.~N. and {Chalumeau}, A. and {Champion}, D.~J. and {Chanlaridis}, S. and {Chen}, S. and {Cognard}, I. and {Dandapat}, S. and {Deb}, D. and {Desai}, S. and {Desvignes}, G. and {Dhanda-Batra}, N. and {Dwivedi}, C. and {Falxa}, M. and {Ferdman}, R.~D. and {Franchini}, A. and {Gair}, J.~R. and {Goncharov}, B. and {Gopakumar}, A. and {Graikou}, E. and {Grie{\ss}meier}, J. -M. and {Guillemot}, L. and {Guo}, Y.~J. and {Gupta}, Y. and {Hisano}, S. and {Hu}, H. and {Iraci}, F. and {Izquierdo-Villalba}, D. and {Jang}, J. and {Jawor}, J. and {Janssen}, G.~H. and {Jessner}, A. and {Joshi}, B.~C. and {Kareem}, F. and {Karuppusamy}, R. and {Keane}, E.~F. and {Keith}, M.~J. and {Kharbanda}, D. and {Kikunaga}, T. and {Kolhe}, N. and {Kramer}, M. and {Krishnakumar}, M.~A. and {Lackeos}, K. and {Lee}, K.~J. and {Liu}, K. and {Liu}, Y. and {Lyne}, A.~G. and {McKee}, J.~W. and {Maan}, Y. and {Main}, R.~A. and {Mickaliger}, M.~B. and {Ni{\c{t}}u}, I.~C. and {Nobleson}, K. and {Paladi}, A.~K. and {Parthasarathy}, A. and {Perera}, B.~B.~P. and {Perrodin}, D. and {Petiteau}, A. and {Porayko}, N.~K. and {Possenti}, A. and {Prabu}, T. and {Quelquejay Leclere}, H. and {Rana}, P. and {Samajdar}, A. and {Sanidas}, S.~A. and {Sesana}, A. and {Shaifullah}, G. and {Singha}, J. and {Speri}, L. and {Spiewak}, R. and {Srivastava}, A. and {Stappers}, B.~W. and {Surnis}, M. and {Susarla}, S.~C. and {Susobhanan}, A. and {Takahashi}, K. and {Tarafdar}, P. and {Theureau}, G. and {Tiburzi}, C. and {van der Wateren}, E. and {Vecchio}, A. and {Venkatraman Krishnan}, V. and {Verbiest}, J.~P.~W. and {Wang}, J. and {Wang}, L. and {Wu}, Z.},
        title = "{The second data release from the European Pulsar Timing Array. III. Search for gravitational wave signals}",
      journal = {\aap},
         year = 2023,
        month = oct,
       volume = {678},
          eid = {A50},
        pages = {A50},
          doi = {10.1051/0004-6361/202346844},
archivePrefix = {arXiv},
       eprint = {2306.16214},
 primaryClass = {astro-ph.HE},
       adsurl = {https://ui.adsabs.harvard.edu/abs/2023A&A...678A..50E}
}

@ARTICLE{Reardon2023,
       author = {{Reardon}, Daniel J. and {Zic}, Andrew and {Shannon}, Ryan M. and {Hobbs}, George B. and {Bailes}, Matthew and {Di Marco}, Valentina and {Kapur}, Agastya and {Rogers}, Axl F. and {Thrane}, Eric and {Askew}, Jacob and {Bhat}, N.~D. Ramesh and {Cameron}, Andrew and {Cury{\l}o}, Ma{\l}gorzata and {Coles}, William A. and {Dai}, Shi and {Goncharov}, Boris and {Kerr}, Matthew and {Kulkarni}, Atharva and {Levin}, Yuri and {Lower}, Marcus E. and {Manchester}, Richard N. and {Mandow}, Rami and {Miles}, Matthew T. and {Nathan}, Rowina S. and {Os{\l}owski}, Stefan and {Russell}, Christopher J. and {Spiewak}, Ren{\'e}e and {Zhang}, Songbo and {Zhu}, Xing-Jiang},
        title = "{Search for an Isotropic Gravitational-wave Background with the Parkes Pulsar Timing Array}",
      journal = {\apjl},
         year = 2023,
        month = jul,
       volume = {951},
       number = {1},
          eid = {L6},
        pages = {L6},
          doi = {10.3847/2041-8213/acdd02},
archivePrefix = {arXiv},
       eprint = {2306.16215},
 primaryClass = {astro-ph.HE},
       adsurl = {https://ui.adsabs.harvard.edu/abs/2023ApJ...951L...6R}
}

@ARTICLE{Shu2020,
       author = {{Suh}, Hyewon and {Civano}, Francesca and {Trakhtenbrot}, Benny and {Shankar}, Francesco and {Hasinger}, G{\"u}nther and {Sanders}, David B. and {Allevato}, Viola},
        title = "{No Significant Evolution of Relations between Black Hole Mass and Galaxy Total Stellar Mass Up to z {\ensuremath{\sim}} 2.5}",
      journal = {\apj},
         year = 2020,
        month = jan,
       volume = {889},
       number = {1},
          eid = {32},
        pages = {32},
          doi = {10.3847/1538-4357/ab5f5f},
archivePrefix = {arXiv},
       eprint = {1912.02824},
 primaryClass = {astro-ph.GA},
       adsurl = {https://ui.adsabs.harvard.edu/abs/2020ApJ...889...32S}
}

@ARTICLE{Bonoli2025,
       author = {{Bonoli}, Silvia and {Izquierdo-Villalba}, David and {Spinoso}, Daniele and {Colpi}, Monica and {Sesana}, Alberto and {Polkas}, Markos and {Springel}, Volker},
        title = "{Constraints on the early growth of massive black holes from PTA and JWST with L-GalaxiesBH}",
      journal = {arXiv e-prints},
         year = 2025,
        month = sep,
          eid = {arXiv:2509.12325},
        pages = {arXiv:2509.12325},
archivePrefix = {arXiv},
       eprint = {2509.12325},
 primaryClass = {astro-ph.GA},
       adsurl = {https://ui.adsabs.harvard.edu/abs/2025arXiv250912325B}
}

@ARTICLE{DongPaez2025,
       author = {{Dong-P{\'a}ez}, Chi An and {Volonteri}, Marta and {Dubois}, Yohan and {Beckmann}, Ricarda S. and {Trebitsch}, Maxime},
        title = "{Wandering and escaping: Recoiling massive black holes in cosmological simulations}",
      journal = {\aap},
         year = 2025,
        month = mar,
       volume = {695},
          eid = {A231},
        pages = {A231},
          doi = {10.1051/0004-6361/202453070},
archivePrefix = {arXiv},
       eprint = {2412.02374},
 primaryClass = {astro-ph.GA},
       adsurl = {https://ui.adsabs.harvard.edu/abs/2025A&A...695A.231D}
}

@ARTICLE{Huang2018,
       author = {{Huang}, Kuan-Wei and {Di Matteo}, Tiziana and {Bhowmick}, Aklant K. and {Feng}, Yu and {Ma}, Chung-Pei},
        title = "{BLUETIDES simulation: establishing black hole-galaxy relations at high redshift}",
      journal = {\mnras},
         year = 2018,
        month = aug,
       volume = {478},
       number = {4},
        pages = {5063-5073},
          doi = {10.1093/mnras/sty1329},
archivePrefix = {arXiv},
       eprint = {1801.04951},
 primaryClass = {astro-ph.GA},
       adsurl = {https://ui.adsabs.harvard.edu/abs/2018MNRAS.478.5063H}
}

@ARTICLE{Volonteri2016,
       author = {{Volonteri}, M. and {Dubois}, Y. and {Pichon}, C. and {Devriendt}, J.},
        title = "{The cosmic evolution of massive black holes in the Horizon-AGN simulation}",
      journal = {\mnras},
         year = 2016,
        month = aug,
       volume = {460},
       number = {3},
        pages = {2979-2996},
          doi = {10.1093/mnras/stw1123},
archivePrefix = {arXiv},
       eprint = {1602.01941},
 primaryClass = {astro-ph.GA},
       adsurl = {https://ui.adsabs.harvard.edu/abs/2016MNRAS.460.2979V}
}

@ARTICLE{Habouzit2019,
       author = {{Habouzit}, M{\'e}lanie and {Genel}, Shy and {Somerville}, Rachel S. and {Kocevski}, Dale and {Hirschmann}, Michaela and {Dekel}, Avishai and {Choi}, Ena and {Nelson}, Dylan and {Pillepich}, Annalisa and {Torrey}, Paul and {Hernquist}, Lars and {Vogelsberger}, Mark and {Weinberger}, Rainer and {Springel}, Volker},
        title = "{Linking galaxy structural properties and star formation activity to black hole activity with IllustrisTNG}",
      journal = {\mnras},
         year = 2019,
        month = apr,
       volume = {484},
       number = {4},
        pages = {4413-4443},
          doi = {10.1093/mnras/stz102},
archivePrefix = {arXiv},
       eprint = {1809.05588},
 primaryClass = {astro-ph.GA},
       adsurl = {https://ui.adsabs.harvard.edu/abs/2019MNRAS.484.4413H}
}

@ARTICLE{Hu2008,
       author = {{Hu}, Jian},
        title = "{The black hole mass-stellar velocity dispersion correlation: bulges versus pseudo-bulges}",
      journal = {\mnras},
         year = 2008,
        month = jun,
       volume = {386},
       number = {4},
        pages = {2242-2252},
          doi = {10.1111/j.1365-2966.2008.13195.x},
archivePrefix = {arXiv},
       eprint = {0801.1481},
 primaryClass = {astro-ph},
       adsurl = {https://ui.adsabs.harvard.edu/abs/2008MNRAS.386.2242H}
}

@ARTICLE{GrahamANDLi2009,
       author = {{Graham}, Alister W. and {Li}, I.-hui},
        title = "{The M $_{bh}$-{\ensuremath{\sigma}} Diagram and the Offset Nature of Barred Active Galaxies}",
      journal = {\apj},
         year = 2009,
        month = jun,
       volume = {698},
       number = {1},
        pages = {812-818},
          doi = {10.1088/0004-637X/698/1/812},
archivePrefix = {arXiv},
       eprint = {0904.1290},
 primaryClass = {astro-ph.CO},
       adsurl = {https://ui.adsabs.harvard.edu/abs/2009ApJ...698..812G}
}

@ARTICLE{Weller2023,
       author = {{Weller}, Emma Jane and {Pacucci}, Fabio and {Natarajan}, Priyamvada and {Di Matteo}, Tiziana},
        title = "{Overmassive central black holes in the cosmological simulations ASTRID and Illustris TNG50}",
      journal = {\mnras},
         year = 2023,
        month = jul,
       volume = {522},
       number = {4},
        pages = {4963-4971},
          doi = {10.1093/mnras/stad1362},
archivePrefix = {arXiv},
       eprint = {2305.02335},
 primaryClass = {astro-ph.GA},
       adsurl = {https://ui.adsabs.harvard.edu/abs/2023MNRAS.522.4963W}
}

@ARTICLE{Ferre_Mateu2021,
       author = {{Ferr{\'e}-Mateu}, A. and {Mezcua}, M. and {Barrows}, R.~S.},
        title = "{A search for active galactic nuclei in low-mass compact galaxies}",
      journal = {\mnras},
         year = 2021,
        month = oct,
       volume = {506},
       number = {4},
        pages = {4702-4714},
          doi = {10.1093/mnras/stab1915},
archivePrefix = {arXiv},
       eprint = {2107.02141},
 primaryClass = {astro-ph.GA},
       adsurl = {https://ui.adsabs.harvard.edu/abs/2021MNRAS.506.4702F}
}

@ARTICLE{Lu2025,
       author = {{Lu}, Yuhao and {SiTu}, HengJian and {Li}, Jie and {Li}, Yixuan and {Liu}, Yang and {Lin}, Wenbin and {Wang}, Yu},
        title = "{287,872 Supermassive Black Holes Masses: Deep Learning Approaching Reverberation Mapping Accuracy}",
      journal = {arXiv e-prints},
         year = 2025,
        month = dec,
          eid = {arXiv:2512.04803},
        pages = {arXiv:2512.04803},
          doi = {10.48550/arXiv.2512.04803},
archivePrefix = {arXiv},
       eprint = {2512.04803},
 primaryClass = {astro-ph.GA},
       adsurl = {https://ui.adsabs.harvard.edu/abs/2025arXiv251204803L}
}

@ARTICLE{vanDokkum2025,
       author = {{van Dokkum}, Pieter and {Jennings}, Connor and {Pasha}, Imad and {Conroy}, Charlie and {Kaul}, Ish and {Abraham}, Roberto and {Danieli}, Shany and {Romanowsky}, Aaron J. and {Tremblay}, Grant},
        title = "{JWST Confirmation of a Runaway Supermassive Black Hole via its Supersonic Bow Shock}",
      journal = {arXiv e-prints},
         year = 2025,
        month = dec,
          eid = {arXiv:2512.04166},
        pages = {arXiv:2512.04166},
          doi = {10.48550/arXiv.2512.04166},
archivePrefix = {arXiv},
       eprint = {2512.04166},
 primaryClass = {astro-ph.GA},
       adsurl = {https://ui.adsabs.harvard.edu/abs/2025arXiv251204166V}
}

@ARTICLE{vanDokkum2023,
       author = {{van Dokkum}, Pieter and {Pasha}, Imad and {Buzzo}, Maria Luisa and {LaMassa}, Stephanie and {Shen}, Zili and {Keim}, Michael A. and {Abraham}, Roberto and {Conroy}, Charlie and {Danieli}, Shany and {Mitra}, Kaustav and {Nagai}, Daisuke and {Natarajan}, Priyamvada and {Romanowsky}, Aaron J. and {Tremblay}, Grant and {Urry}, C. Megan and {van den Bosch}, Frank C.},
        title = "{A Candidate Runaway Supermassive Black Hole Identified by Shocks and Star Formation in its Wake}",
      journal = {\apjl},
         year = 2023,
        month = apr,
       volume = {946},
       number = {2},
          eid = {L50},
        pages = {L50},
          doi = {10.3847/2041-8213/acba86},
archivePrefix = {arXiv},
       eprint = {2302.04888},
 primaryClass = {astro-ph.GA},
       adsurl = {https://ui.adsabs.harvard.edu/abs/2023ApJ...946L..50V}
}

@ARTICLE{Barrows2025,
       author = {{Barrows}, R. Scott and {Comerford}, Julia M. and {Negus}, James and {Muller-Sanchez}, Francisco},
        title = "{Recoiling Black Hole Candidates from Spatially Offset Broad Emission Lines in MaNGA}",
      journal = {\apj},
         year = 2025,
        month = oct,
       volume = {992},
       number = {1},
          eid = {38},
        pages = {38},
          doi = {10.3847/1538-4357/adfbf3},
archivePrefix = {arXiv},
       eprint = {2512.00174},
 primaryClass = {astro-ph.GA},
       adsurl = {https://ui.adsabs.harvard.edu/abs/2025ApJ...992...38B}
}

@ARTICLE{Islam2026,
       author = {{Islam}, Tousif and {Venumadhav}, Tejaswi and {Wadekar}, Digvijay},
        title = "{Progenitor of the recoiling super-massive black hole RBH-1 identified using HST/JWST imaging}",
      journal = {arXiv e-prints},
         year = 2026,
        month = jan,
          eid = {arXiv:2601.18986},
        pages = {arXiv:2601.18986},
          doi = {10.48550/arXiv.2601.18986},
archivePrefix = {arXiv},
       eprint = {2601.18986},
 primaryClass = {astro-ph.HE},
       adsurl = {https://ui.adsabs.harvard.edu/abs/2026arXiv260118986I}
}

@ARTICLE{SanchezAlmeida2023,
       author = {{S{\'a}nchez Almeida}, J.},
        title = "{Supermassive black hole wake or bulgeless edge-on galaxy?. II. Order-of-magnitude analysis of the two physical scenarios}",
      journal = {\aap},
         year = 2023,
        month = oct,
       volume = {678},
          eid = {A118},
        pages = {A118},
          doi = {10.1051/0004-6361/202347098},
archivePrefix = {arXiv},
       eprint = {2309.02494},
 primaryClass = {astro-ph.GA},
       adsurl = {https://ui.adsabs.harvard.edu/abs/2023A&A...678A.118S}
}

@ARTICLE{Ramsden2026,
       author = {{Ramsden}, Paige and {McGee}, Sean L. and {Nicholl}, Matt},
        title = "{An Unexpected Population of Quenched Galaxies Harboring Undermassive SMBHs Revealed by Tidal Disruption Events}",
      journal = {\apjl},
         year = 2026,
        month = feb,
       volume = {998},
       number = {1},
          eid = {L25},
        pages = {L25},
          doi = {10.3847/2041-8213/ae3daa},
archivePrefix = {arXiv},
       eprint = {2601.20519},
 primaryClass = {astro-ph.GA},
       adsurl = {https://ui.adsabs.harvard.edu/abs/2026ApJ...998L..25R}
}

@ARTICLE{Freeman1970,
       author = {{Freeman}, K.~C.},
        title = "{On the Disks of Spiral and S0 Galaxies}",
      journal = {\apj},
         year = 1970,
        month = jun,
       volume = {160},
        pages = {811},
          doi = {10.1086/150474},
       adsurl = {https://ui.adsabs.harvard.edu/abs/1970ApJ...160..811F}
}

@ARTICLE{Kocevski2023,
       author = {{Kocevski}, Dale D. and {Onoue}, Masafusa and {Inayoshi}, Kohei and {Trump}, Jonathan R. and {Arrabal Haro}, Pablo and {Grazian}, Andrea and {Dickinson}, Mark and {Finkelstein}, Steven L. and {Kartaltepe}, Jeyhan S. and {Hirschmann}, Michaela and {Aird}, James and {Holwerda}, Benne W. and {Fujimoto}, Seiji and {Juneau}, St{\'e}phanie and {Amor{\'\i}n}, Ricardo O. and {Backhaus}, Bren E. and {Bagley}, Micaela B. and {Barro}, Guillermo and {Bell}, Eric F. and {Bisigello}, Laura and {Calabr{\`o}}, Antonello and {Cleri}, Nikko J. and {Cooper}, M.~C. and {Ding}, Xuheng and {Grogin}, Norman A. and {Ho}, Luis C. and {Hutchison}, Taylor A. and {Inoue}, Akio K. and {Jiang}, Linhua and {Jones}, Brenda and {Koekemoer}, Anton M. and {Li}, Wenxiu and {Li}, Zhengrong and {McGrath}, Elizabeth J. and {Molina}, Juan and {Papovich}, Casey and {P{\'e}rez-Gonz{\'a}lez}, Pablo G. and {Pirzkal}, Nor and {Wilkins}, Stephen M. and {Yang}, Guang and {Yung}, L.~Y. Aaron},
        title = "{Hidden Little Monsters: Spectroscopic Identification of Low-mass, Broad-line AGNs at z > 5 with CEERS}",
      journal = {\apjl},
         year = 2023,
        month = sep,
       volume = {954},
       number = {1},
          eid = {L4},
        pages = {L4},
          doi = {10.3847/2041-8213/ace5a0},
archivePrefix = {arXiv},
       eprint = {2302.00012},
 primaryClass = {astro-ph.GA},
       adsurl = {https://ui.adsabs.harvard.edu/abs/2023ApJ...954L...4K}
}

@ARTICLE{Jones2025,
       author = {{Jones}, Brenda L. and {Kocevski}, Dale D. and {Pacucci}, Fabio and {Taylor}, Anthony J. and {Finkelstein}, Steven L. and {Buchner}, Johannes and {Trump}, Jonathan R. and {Somerville}, Rachel S. and {Hirschmann}, Michaela and {Yung}, L.~Y. Aaron and {Barro}, Guillermo and {Bell}, Eric F. and {Bisigello}, Laura and {Calabro}, Antonello and {Cleri}, Nikko J. and {Dekel}, Avishai and {Dickinson}, Mark and {Gandolfi}, Giovanni and {Giavalisco}, Mauro and {Grogin}, Norman A. and {Inayoshi}, Kohei and {Kartaltepe}, Jeyhan S. and {Koekemoer}, Anton M. and {Napolitano}, Lorenzo and {Onoue}, Masafusa and {Ravindranath}, Swara and {Rodighiero}, Giulia and {Wilkins}, Stephen M.},
        title = "{The $M_{\rm BH}-M_{*}$ Relationship at $3<z<7$: Big Black Holes in Little Red Dots}",
      journal = {arXiv e-prints},
         year = 2025,
        month = oct,
          eid = {arXiv:2510.07376},
        pages = {arXiv:2510.07376},
          doi = {10.48550/arXiv.2510.07376},
archivePrefix = {arXiv},
       eprint = {2510.07376},
 primaryClass = {astro-ph.GA},
       adsurl = {https://ui.adsabs.harvard.edu/abs/2025arXiv251007376J}
}

@ARTICLE{Pacucci2023,
       author = {{Pacucci}, Fabio and {Nguyen}, Bao and {Carniani}, Stefano and {Maiolino}, Roberto and {Fan}, Xiaohui},
        title = "{JWST CEERS and JADES Active Galaxies at z = 4-7 Violate the Local M $_{{\textbullet}}$-M $_{{\ensuremath{\star}}}$ Relation at >3{\ensuremath{\sigma}}: Implications for Low-mass Black Holes and Seeding Models}",
      journal = {\apjl},
         year = 2023,
        month = nov,
       volume = {957},
       number = {1},
          eid = {L3},
        pages = {L3},
          doi = {10.3847/2041-8213/ad0158},
archivePrefix = {arXiv},
       eprint = {2308.12331},
 primaryClass = {astro-ph.GA},
       adsurl = {https://ui.adsabs.harvard.edu/abs/2023ApJ...957L...3P}
}

@ARTICLE{Dave2019,
       author = {{Dav{\'e}}, Romeel and {Angl{\'e}s-Alc{\'a}zar}, Daniel and {Narayanan}, Desika and {Li}, Qi and {Rafieferantsoa}, Mika H. and {Appleby}, Sarah},
        title = "{SIMBA: Cosmological simulations with black hole growth and feedback}",
      journal = {\mnras},
         year = 2019,
        month = jun,
       volume = {486},
       number = {2},
        pages = {2827-2849},
          doi = {10.1093/mnras/stz937},
archivePrefix = {arXiv},
       eprint = {1901.10203},
 primaryClass = {astro-ph.GA},
       adsurl = {https://ui.adsabs.harvard.edu/abs/2019MNRAS.486.2827D}
}

@ARTICLE{Bird2022,
       author = {{Bird}, Simeon and {Ni}, Yueying and {Di Matteo}, Tiziana and {Croft}, Rupert and {Feng}, Yu and {Chen}, Nianyi},
        title = "{The ASTRID simulation: galaxy formation and reionization}",
      journal = {\mnras},
         year = 2022,
        month = may,
       volume = {512},
       number = {3},
        pages = {3703-3716},
          doi = {10.1093/mnras/stac648},
archivePrefix = {arXiv},
       eprint = {2111.01160},
 primaryClass = {astro-ph.GA},
       adsurl = {https://ui.adsabs.harvard.edu/abs/2022MNRAS.512.3703B}
}

@ARTICLE{Nelson2019,
	author = {{Nelson}, Dylan and {Pillepich}, Annalisa and {Springel}, Volker and
	{Pakmor}, R{\"u}diger and {Weinberger}, Rainer and {Genel}, Shy and
	{Torrey}, Paul and {Vogelsberger}, Mark and {Marinacci}, Federico and
	{Hernquist}, Lars},
	title = "{First results from the TNG50 simulation: galactic outflows driven by supernovae and black hole feedback}",
	journal = {\mnras},
	year = 2019,
	month = dec,
	volume = {490},
	number = {3},
	pages = {3234-3261},
	doi = {10.1093/mnras/stz2306},
	archivePrefix = {arXiv},
	eprint = {1902.05554},
	primaryClass = {astro-ph.GA},
	adsurl = {https://ui.adsabs.harvard.edu/abs/2019MNRAS.490.3234N}
}

@ARTICLE{Pillepich2019,
	author = {{Pillepich}, Annalisa and {Nelson}, Dylan and {Springel}, Volker and
	{Pakmor}, R{\"u}diger and {Torrey}, Paul and {Weinberger}, Rainer and
	{Vogelsberger}, Mark and {Marinacci}, Federico and {Genel}, Shy and
	{van der Wel}, Arjen and {Hernquist}, Lars},
	title = "{First results from the TNG50 simulation: the evolution of stellar and gaseous discs across cosmic time}",
	journal = {\mnras},
	year = 2019,
	month = dec,
	volume = {490},
	number = {3},
	pages = {3196-3233},
	doi = {10.1093/mnras/stz2338},
	archivePrefix = {arXiv},
	eprint = {1902.05553},
	primaryClass = {astro-ph.GA},
	adsurl = {https://ui.adsabs.harvard.edu/abs/2019MNRAS.490.3196P}
}

@ARTICLE{Herrero-Carrion2026,
       author = {{Herrero-Carri{\'o}n}, Diego and {Spinoso}, Daniele and {Izquierdo-Villalba}, David and {Su}, Tong and {Bonoli}, Silvia and {Renard}, Pablo},
        title = "{Back to basics: Little Red Dots as galaxies and dust-obscured AGNs in a synthetic NIRCam sky simulated with L-GalaxiesBH}",
      journal = {\mnras},
         year = 2026,
        month = mar,
          doi = {10.1093/mnras/stag478},
archivePrefix = {arXiv},
       eprint = {2511.10725},
 primaryClass = {astro-ph.GA},
       adsurl = {https://ui.adsabs.harvard.edu/abs/2026MNRAS.tmp..450H}
}

\begin{appendix}
\section{Comparison of the $M_{\rm BH}/M_*$ ratio with Hydrodynamical Simulations}\label{Appendix:Comparison_MBH_Mstellar}
 In this appendix we compare the evolution of the $M_{BH}/M_*$ ratio (for $M_*\,{>}\,10^{10} M_{\odot}$ galaxies) predicted by \lgbh{} with that obtained from five cosmological hydrodynamical simulations: \texttt{TNG-50}, \texttt{TNG-100}, \texttt{TNG-300} \citep{NelsonTNGDataReleas2019,Nelson2019,Pillepich2019}, \texttt{SIMBA} \citep[Flagship,][]{Dave2019}, and \texttt{Astrid} \citep{Bird2022}. These simulations differ in their galaxy formation prescriptions, numerical resolution, and adopted initial black hole seed masses.\\

As shown in Fig.~\ref{fig:M_BH_M_gal_Evolution_Comparison}, the largest differences between \lgbh{} and the hydrodynamical simulations appear at $z\,{>}\,3$, where the latter predict typical $M_{BH}/M_*$ ratios between a factor of ${\sim}\,1\,{-}\,3$ dex larger than those obtained with \lgbh{}. This discrepancy is primarily driven by the assumed seed masses. In \lgbh{}, MBHs are predominantly seeded through a light-seed channel associated with PopIII remnants, with typical initial masses of $10\,{-}\,100\,M_{\odot}$. In contrast, most hydrodynamical simulations adopt seed masses of $10^4\,{-}\,10^5\,\msun$, which naturally facilitates the formation of systems with higher $M_{\rm BH}/M_*$ ratios at early times. To enable a more direct comparison and reduce the impact of the seed-mass assumption, we recomputed the $M_{BH}/M_*$ evolution in \lgbh{} considering only systems hosting MBHs with masses ${>}\,10^6 \msun$, comparable to the effective minimum black hole mass present in hydrodynamical simulations after seeding. Under this selection, the predicted $M_{\rm BH}/M_*$ ratios in \lgbh{} shift toward higher values and become more consistent with those obtained from the hydrodynamical simulations. The remaining differences are likely driven by other aspects of the galaxy and MBH assembly models, such as the treatment of supernova and MBH feedback or star formation. At $z\,{<}\,3$, \lgbh{} shows good agreement with the predictions of \texttt{SIMBA} and \texttt{Astrid}, while the \texttt{TNG} simulations still yield $M_{\rm BH}/M_*$ ratios larger by approximately a factor of seven. Nevertheless, regardless of the specific hydrodynamical simulation considered, all models exhibit a qualitatively similar trend to that found in \lgbh{}: the $M_{BH}/M_*$ ratio increases toward lower redshift, although the amplitude of this evolution varies among simulations.

\begin{figure}
    \centering
    \includegraphics[width=1.0\columnwidth]{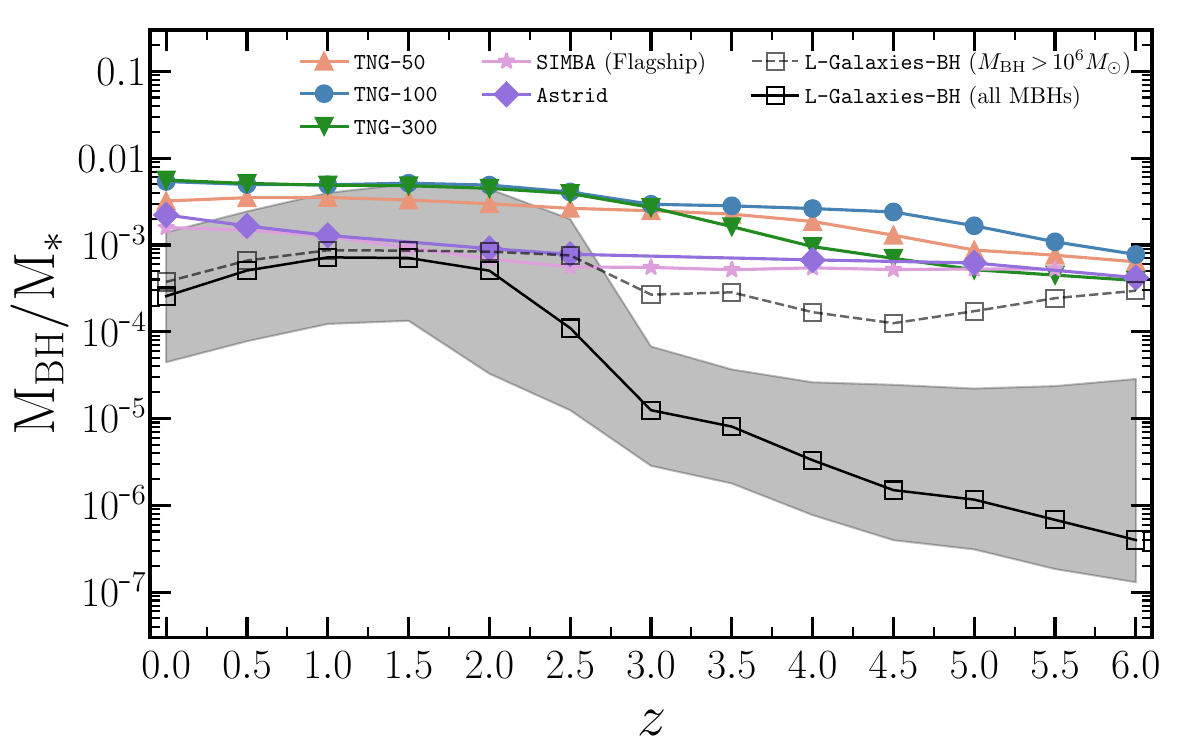}
    \caption{\footnotesize Redshift evolution of the median $M_{\rm BH}/M_*$ ratio for galaxies with  $M_*\,{>}\,10^{10} \, M_{\odot}$ in \texttt{TNG-50}, \texttt{TNG-100}, \texttt{TNG-300}, \texttt{SIMBA} (flagship), and \texttt{Astrid} hydrodynamical simulations.The results are compared with \lgbh{} accounting for all the galaxies hosting a MBH (solid line) and galaxies hosting only MBHs with $M_{\rm BH}\,{>}\,10^6 \rm M_{\odot}$. The shaded areas correspond to the $16^{\rm th}\,{-}\,84^{\rm th}$ percentiles.} 
    \label{fig:M_BH_M_gal_Evolution_Comparison}
\end{figure}

\section{The duty cycle of the average, overmassive and unedermassive population}\label{Appendix:DutyCycle}
In this appendix, we quantify the degree of desynchronization between the growth of galaxies and their central MBHs in the undermassive population. To this end, Fig.~\ref{fig:Redshift_Evol_DutyCicle} presents the redshift evolution of the MBH duty cycle for the different samples. We define the duty cycle as the fraction of time since MBH formation during which the MBH is actively accreting (including super-Eddington and Eddington phases). The median population (\Onep) exhibits a clear redshift dependence: the duty cycle is relatively high (${\sim}\, 30 \,{-}\,40\%$) at $z \,{>}\, 3$, while it declines to $\sim 10 \,{-}\, 20\%$ at lower redshifts. A similar evolutionary trend is found for the \Twop{} and \Threep{} samples, although with systematically higher duty cycles at all redshifts, reaching values of ${\sim}\, 60\,{-}\,70\%$ at $z \,{>}\, 3$. In contrast, the undermassive samples (\Twom{} and \Threem{}) show little to no redshift evolution, maintaining low duty cycles of ${\sim}\, 5 \,{-}\, 20\%$ across all epochs. This difference is particularly pronounced at stellar masses $M_* \,{<}\, 10^{10}\,\msun$, where, at fixed stellar mass, the duty cycles of the undermassive systems are lower by a factor of ${\sim}\, 4 \,{-}\,5$ compared to the median population.\\

Taken together, these trends provide a quantitative measure of the degree of desynchronization between galaxy growth and MBH accretion in the undermassive population. At fixed stellar mass, MBHs in such a sample spend substantially less time in active accretion phases than their counterparts in the median population. This indicates that, although their host galaxies continue to build up stellar mass, MBH growth is significantly delayed relative to the general population. Consequently, these MBHs remain underdeveloped and follow evolutionary pathways that are largely decoupled from the canonical co-evolutionary track.

\begin{figure}
    \centering
    \includegraphics[width=0.75\columnwidth]{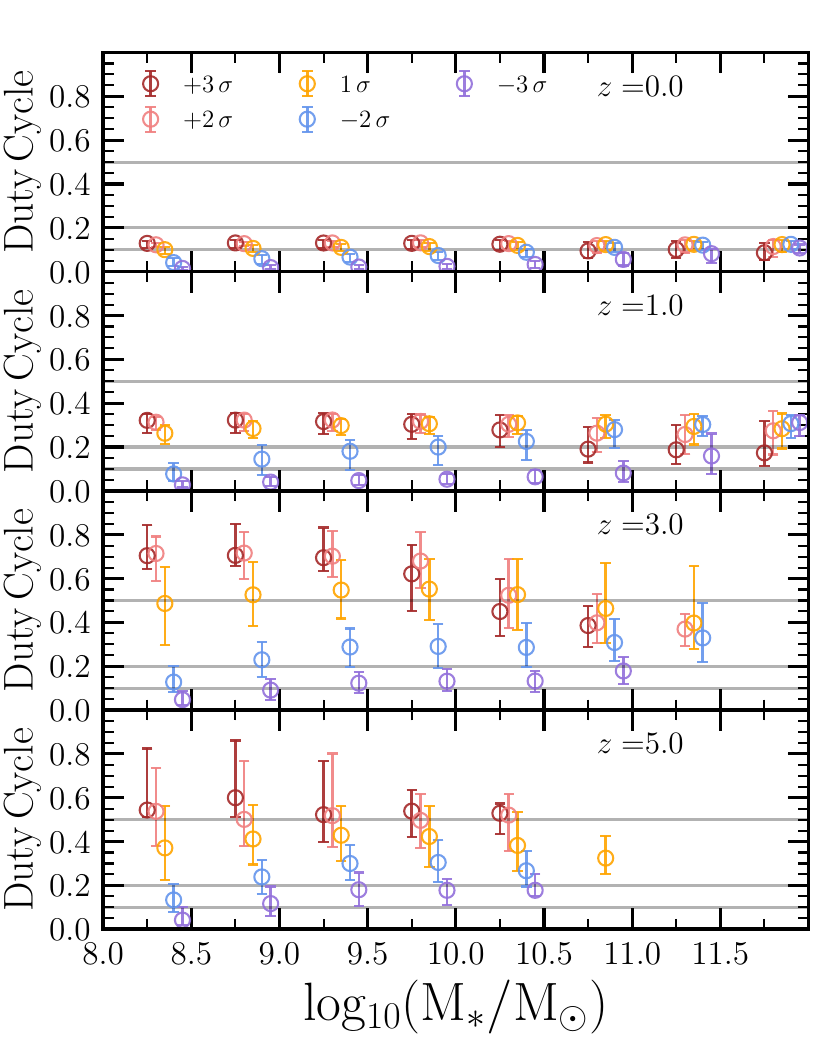}
    \caption{\footnotesize Duty cycle as a function of stellar mass for the different samples. Each row corresponds to a given redshift bin (from top to bottom: $z \,{=}\, 0, 1, 3, 5$). Points indicate median values, while error bars show the $16^{\rm th}\,{-}\,84^{\rm th}$ percentiles. For clarity, the different samples are slightly offset along the $M_*$ axis (by 0.05 dex) to avoid overlap.} 
    \label{fig:Redshift_Evol_DutyCicle}
\end{figure}

\end{appendix}
\end{document}